\documentclass[twocolumn]{revtex4-2}
\usepackage[english]{babel}
\usepackage{amsmath}
\usepackage{amssymb}
\usepackage{graphicx}
\usepackage{rotating}
\usepackage{float}

\begin{document}

\title{Chiral magnetic domain walls under transverse fields: a semi-analytical model}
\date{\today}

\author{Pierre G{\'{e}}hanne}

\author{Andr{\'{e}} Thiaville}
\email[]{andre.thiaville@universite-paris-saclay.fr}

\author{Stanislas Rohart}

\author{Vincent Jeudy}

\affiliation{Laboratoire de Physique des Solides, Universit{\'{e}} Paris-Saclay, CNRS UMR 8502, 91405 Orsay, France}

\begin{abstract}
An analytical model for the domain wall structure in ultrathin films with perpendicular easy axis
and interfacial Dzyaloshinskii-Moriya interaction, submitted to an arbitrary in-plane magnetic field, 
is presented.
Its solution is simplified to the numerical minimization of an analytic function of just
one variable.
The model predictions are compared to numerical micromagnetic simulations, using parameters
of existing samples, revealing a very good agreement.
Remaining differences are analyzed, and partly corrected.
Differences with the predictions of the simplest model, usually found in the literature, in which only the 
domain wall moment's in-plane orientation can vary, are exemplified.
The model allows accurate computations, as a function of in-plane field module and orientation, of the 
domain wall tension and width, quantities controlling the creep motion of domain walls in such films.
\end{abstract}

\maketitle


\section{Introduction}

The interfacial Dzyaloshinskii-Moriya (DMI) interaction was shown, in the last years, to have an 
important role on the magnetization statics and dynamics \cite{Hellman17}, especially 
in the case of magnetic domain walls.
This holds not only for epitaxial atomic monolayers on single-crystal substrates \cite{Bode07,Heide08},
but also for the polycrystalline ultrathin films which are at the heart of present spintronic devices
\cite{Thiaville12}.
The interfacial DMI, like the DMI introduced many years ago \cite{Dzialoshinskii57,Moriya60}, is 
allowed only when spatial inversion symmetry is broken.
Such symmetry breaking takes place naturally at interfaces \cite{Fert90}.
DMI is expressed as an exchange interaction with an anti-symmetric matrix.
The form of this matrix is dictated by the symmetry of the atomic arrangement, according to
the Moriya rules \cite{Moriya60}.
Interfacial DMI, in the limiting case of maximal symmetry compatible with the presence
of an interface (like for two amorphous materials on each side of the interface), and specializing
to a film with perpendicular magnetization, applies to the moments in such domain walls 
a chiral in-plane effective field.
As a consequence, the application of in-plane fields on such samples has become a very common
experimental tool to study and control the effect of the interfacial DMI.

As the applied fields can be large (because the DMI-induced effective field can also be), the effect of
these fields on the domain wall structure and dynamics should be precisely appreciated.
However, the complete re-calculation of the one-dimensional domain wall profile under an in-plane 
field has not been performed systematically (as the DMI was absent in the previous 
works \cite{Kaczer61,Hubert74}), and various approximations have been recently considered 
in the presence of DMI \cite{Thiaville12,Je13,Emori14,Kim16}.
It is the goal of this paper to describe an accurate semi-analytical method to perform such
calculations, based on the `small circle' Ansatz employed by A.~Hubert a long
time ago \cite{Hubert74}, as we have found that this model describes rather well the situation.
The accuracy of these calculations is indeed required for reliably estimating the domain wall 
surface tension \cite{Pellegren17}, which is numerically very sensitive as
it involves the second derivative of the domain wall energy.
The importance of this parameter, different from the domain wall surface energy, for the domain wall motion 
in the creep regime was realized recently \cite{Pellegren17}.
Moreover, the variations of the domain wall width, which have recently been shown to affect
the pinning of domain walls \cite{Gehanne20}, are also obtained by this model.

Along the paper, semi-analytical results are compared to micromagnetic simulations, using as parameters 
those of several Pt/Co ultrathin films that were studied elsewhere \cite{Gehanne20}.
They are also compared to the simplest model (called `constrained model' hereafter) in which the domain
wall profile is fixed, except for the in-plane angle of the domain wall magnetization 
\cite{Thiaville12,Je13,Emori14,Pellegren17} (note that less constrained variants exist, for example
with a variable domain wall width \cite{Kim16}).
After describing generally the small circle model (Sec.~\ref{sec:pc}) \cite{note-PC}, the case where 
the in-plane field is applied along the domain-wall normal is first treated, as it is the most 
considered configuration (Sec.~\ref{sec:perp}).
Then the general case of a field applied at an arbitrary angle with respect to the domain wall
is treated, by the same method (Sec.~\ref{sec:gene}).
The obtained solutions are used to evaluate the important parameters of the domain
wall, namely its width (several definitions are considered, corresponding to different
physical meanings of this width), and energy.
In the last section, the domain wall surface tension is also evaluated.


\section{Small circle model}
\label{sec:pc}

The reference frame used throughout is set by $x$ the direction of the applied field, 
and $z$ the normal to the film.
The domain wall normal is the $\vec{n}$ direction, with $\vec{m}=(0,0,1)$ representing the unit magnetization
vector in the domain for $n <0$ ($n$ is the abscissa along the $\vec{n}$ direction), far from the wall.
In the presence of an in-plane field, magnitude $H$ (positive by construction), the magnetization 
in the domains rotates from $(0,0,\pm 1)$ to $(h,0,\pm \sqrt{1-h^2})$.
We define $h \equiv H/H_{K0}$ the reduced applied field,
$H_{K0}= 2 K_0/(\mu_0 M_\mathrm{s})$ being the effective anisotropy field
of the sample, with $K_0=K_u - \mu_0 M_\mathrm{s}^2/2$ the effective anisotropy
including the thin film demagnetzing effect for perpendicular magnetization, the uniaxial
anisotropy constant $K_u$ itself consisting of bulk crystalline and interface anisotropy.
The interfacial DMI in the considered samples favors N{\'{e}}el walls, with a chirality
fixed by the sign of the DMI constant.
The samples considered for the numerical evaluations are Pt/Co/Pt, Pt/Co/Au and Au/Co/Pt, 
with a nominal cobalt thickness of 0.9~nm, in which the DMI constant $D$ varies widely.

\begin{figure}
\includegraphics[width= 9 cm]{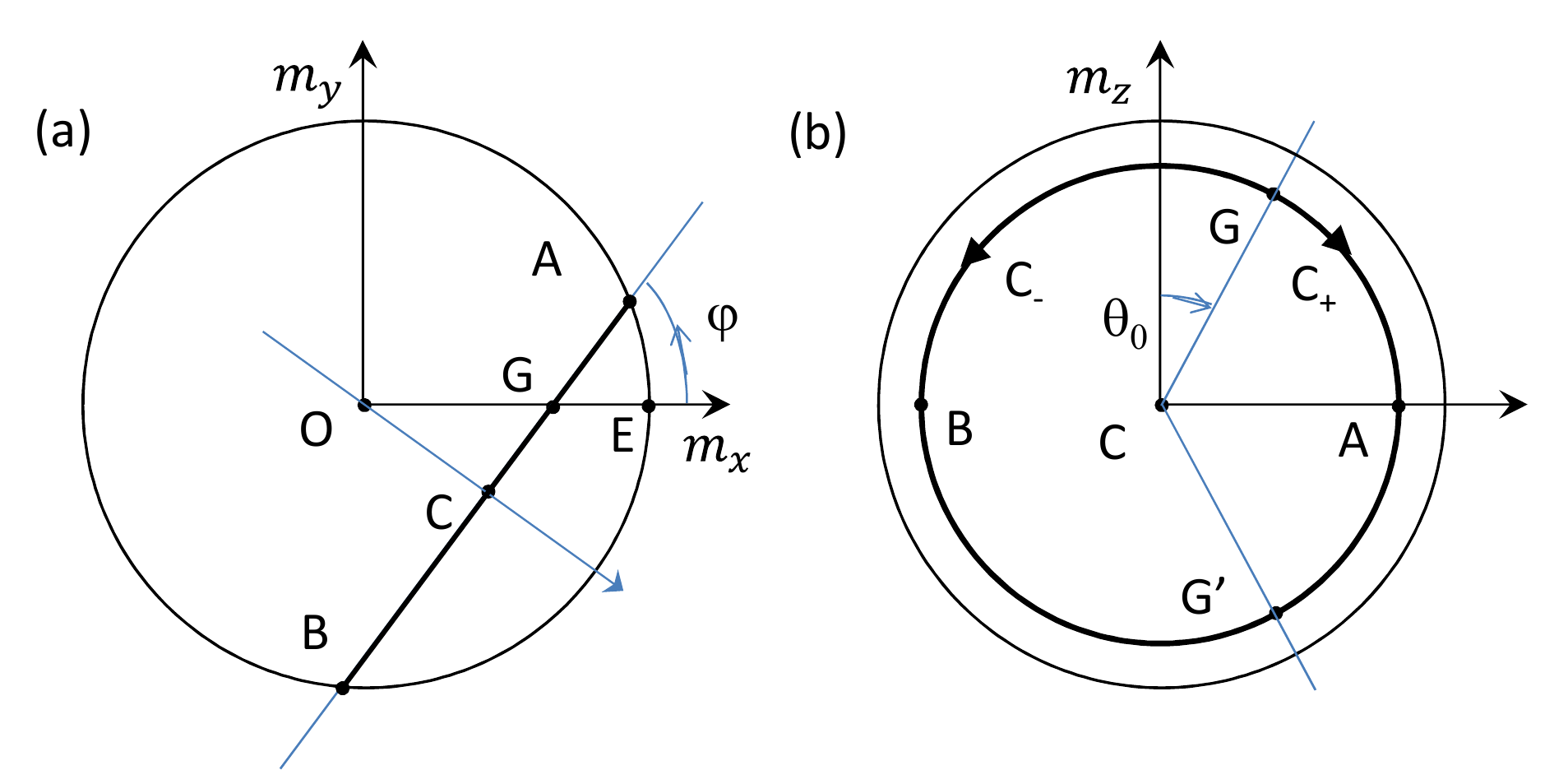}
\caption{
The `small circle' Ansatz.
(a) Top view of the unit sphere for the magnetization vectors, under an in-plane field in the
$+x$ direction. 
Point $G$ depicts the magnetization orientation in the $n<0$ domain (and hides point $G'$ which
corresponds to the $n>0$ domain).
A small circle of the unit sphere, that goes through $G$ and $G'$ and is vertical, is defined 
by the cut angle $\varphi$, which has to be optimized for each situation.
Point $C$ is the center of this small circle.
The in-plane angle of the domain wall magnetization, called $\psi$, is defined by the
angle of OA with OE; it is thus different from $\varphi$.
(b) Side view in the direction normal to the small circle, showing the two paths that connect
$G$ to $G'$, either through $A$ (short path $C_+$ where $\theta$ increases), or through $B$ 
(long path $C_-$ where $\theta$ decreases).
The great circle (radius $1$) is also drawn in thin line.
\label{fig:pc}
}
\end{figure}

The `small circle' Ansatz, is depicted in Fig.~\ref{fig:pc}.
Under the influence of the in-plane field, the magnetization in the domains rotates
out of the poles to points denoted $G$ and $G'$.
A domain wall is, quite generally, a path on the sphere that connects these two points.
The internal magnetostatic energy of the domain wall favors equally the two paths that 
are parallel to the domain wall orientation, whereas DMI favors only one path orthogonal
to the domain wall orientation, and applied field favors also only one path, through
point $E$ of Fig.~\ref{fig:pc}(a).
The idea of the Ansatz is to simplify the task by looking at paths that are contained
within a plane, so moderately long.
This restriction allows analytical calculations nearly up to the end, and
we will show below that it corresponds well to numerical simulation results.
The cut of the unit sphere by a plane gives, by definition, a small circle, hence
the name of this Ansatz.
The family of planes going through points $G$ and $G'$ is described by a single
parameter, the `cut angle' $\varphi$, with $-\pi/2 < \varphi < \pi/2$.
The magnetization position on the small circle is described by an angle $\theta$.
It increases from $\theta_0$ to $\pi - \theta_0$ for the short path $C_+$, and 
decreases from $\theta_0$ to $-\pi -\theta_0$ for the long path $C_-$.
From the drawings, the radius of the small circle is $r=\sqrt{1-h^2 \sin^2 \varphi}$,
and the coordinates of the center $C$ are 
$\overrightarrow{OC}=(h \sin^2 \varphi,-h\sin\varphi \cos\varphi, 0)$.
Thus, the magnetization along the small circle reads
\begin{equation}
\vec{m}= \begin{pmatrix}
h \sin^2 \varphi + r \cos\varphi \sin\theta \\
-h\sin\varphi \cos\varphi +r \sin\varphi \sin\theta \\
r \cos\theta
\end{pmatrix}
\end{equation}
The angle $\theta_0$ along the small circle that corresponds to the domains magnetization
satisfies $\sin\theta_0 = h \cos\varphi / r$ and
$\cos\theta_0=\sqrt{1-h^2}/r$.
Note that by definition one has $0 < \theta_0 < \pi/2$ and $h>0$.


\section{Field normal to the domain wall}
\label{sec:perp}

This is a high symmetry situation (the $x$ axis and the $n$ axis coincide), where the applied field 
$H$ and the DMI-induced effective field at the domain wall are collinear.
When these two fields point in the same sense (and their sum is sufficiently large), 
the solution is the N{\'{e}}el wall of the corresponding chirality. 
When the applied field is opposite to the DMI field and sufficiently large, the solution
is again a N{\'{e}}el wall, with reversed chirality.
When the applied field is close to compensate the DMI effective field, an intermediate
Bloch-N{\'{e}}el wall may appear.

\subsection{Semi-analytical model}
\label{sec:model-normal}

The densities for the exchange, DMI, effective uniaxial anisotropy, Zeeman and domain wall
internal magnetostatic energy are, respectively,
\begin{subequations}
\label{eq:lesE}
\begin{eqnarray}
&\mathcal{E}_\mathrm{exc}& = A r^2 \left( \frac{d\theta}{dn} \right)^2,  \label{subeq:eA} \\
&\mathcal{E}_\mathrm{DMI}& = -D \frac{d\theta}{dn} r ( h \sin^2 \varphi \sin\theta + r \cos\varphi),  
 \label{subeq:eDM} \\
&\mathcal{E}_\mathrm{anis}& = K_0 \left( h^2 \sin^2 \varphi + r^2 \sin^2 \theta \right),  
 \label{subeq:eK} \\
&\mathcal{E}_\mathrm{Z}& = -\mu_0 M_\mathrm{s} H \left( h \sin^2 \varphi + r \cos\varphi \sin\theta \right),  
 \label{subeq:eZ} \\
&\mathcal{E}_\mathrm{BN}& = K \cos^2 \varphi \left( h \cos\varphi - r \sin\theta \right)^2.
 \label{subeq:eBN}
\end{eqnarray}
\end{subequations}
In the last expression, $K$ is the effective magnetostatic term related to the domain wall, i.e. the
magnetostatic cost of a N{\'{e}}el wall.
In the ultrathin limit (sample thickness $t \ll $ domain wall width parameter $\Delta_0$
where $\Delta_0 = \sqrt{A/K_0}$), it reads 
$K \approx \mu_0 M_\mathrm{s}^2 t \ln 2 / (2 \pi \Delta)$ \cite{Tarasenko98}.
The DMI-induced effective field at the domain wall is
$H_\mathrm{DMI}=D/(\mu_0 M_\mathrm{s} \Delta_0)$.

The integral of the total energy density $\mathcal{E}$ has to be minimized with respect to the function 
$\theta(n)$, with the constrains $\theta(-\infty) = \theta_0$, $\theta(+\infty)=\pi-\theta_0$.
Inspection of the terms of Eq.~(\ref{eq:lesE}) shows that (i) the DMI term is the $x$ derivative of some function, 
hence will play no role in the profile $\theta(n)$; (ii) the energy density has the usual expression of domain 
walls, with a gradient squared plus a function of $\theta$.
Therefore, the associated first integral can be written, leading to the angle
differential variation law
\begin{equation}
\label{eq:dtdx}
\frac{d \theta}{d \xi}= \sqrt{1+\kappa \cos^2 \varphi} \left( \sin\theta - \sin\theta_0 \right),
\end{equation}
where the reduced variables $\xi = n/\Delta_0$ and $\kappa = K/K_0$ have been introduced.
This relation can be integrated \cite{Kaczer61}, a surprisingly simple formulation of this result 
being \cite{Hubert74}
\begin{equation}
\label{eq:sint(x)}
\sin\theta = \sin\theta_0 + \frac{\cos^2 \theta_0}{\sin \theta_0 + 
\cosh \left(\xi \cos\theta_0 \sqrt{1+\kappa \cos^2 \varphi} \right)}.
\end{equation}
From Eq.~(\ref{eq:dtdx}), the energy of the domain wall can be evaluated.
It is given by the integral of $\mathcal{E} + K_0 h^2$, the last term having been included to
remove the energy density in the domains.
Using the condition (\ref{eq:dtdx}) to simplify the calculations, one obtains finally
\begin{subequations}
\label{eq:les-sig}
\begin{eqnarray}
\frac{\sigma_+}{\sigma_0} &=& r^2 \sqrt{1+\kappa \cos^2 \varphi} \left[ \cos\theta_0 - 
\left( \frac{\pi}{2}-\theta_0 \right) \sin\theta_0 \right]  \nonumber \\
&-& r \delta \left[ h \sin^2 \varphi \cos\theta_0
+(\frac{\pi}{2}-\theta_0)r \cos\varphi \right] \label{subeq:sig+}\\
\frac{\sigma_-}{\sigma_0} &=& r^2 \sqrt{1+\kappa \cos^2 \varphi} \left[ \cos\theta_0 + 
\left( \frac{\pi}{2}+\theta_0 \right) \sin\theta_0 \right] \nonumber \\
&-& r \delta \left[ h \sin^2 \varphi \cos\theta_0
-(\frac{\pi}{2}+\theta_0)r \cos\varphi \right] \label{subeq:sig-}.
\end{eqnarray}
\end{subequations}
In this expression, $\delta = D/(2\sqrt{A K_0}) = (2/\pi) D/D_c$ is the reduced DMI constant, with
$D_c= (4/\pi) \sqrt{A K_0}$ the well-known critical value of DMI [neglecting the Bloch-N{\'{e}}el anisotropy
energy of Eq.~(\ref{subeq:eBN})] above which the uniform magnetic state is unstable.
In the constrained model \cite{Thiaville12,Je13,Emori14} where only the in-plane angle (called $\psi$) of the 
domain wall magnetization can vary, one simply has to minimize versus $\psi$ the expression
$\sigma(\psi)/\sigma_0 = 1+(\kappa/2) \cos^2\psi -(\pi/2)(\delta+h) \cos\psi$.

Let us now look at a few limiting cases.
When $H=0$, the small circle is a great circle so $r=1$, $\theta_0=0$, and the cut angle $\varphi$
is the angle of the domain wall magnetization ($\varphi=\psi=0$ for N{\'{e}}el walls, 
$\pi/2$ for Bloch walls).
The domain wall energy simplifies to
\begin{equation}
\label{eq:H0}
\frac{\sigma_\pm}{\sigma_0}= \sqrt{1+\kappa \cos^2 \varphi} \mp \frac{\pi}{2} \delta \cos\varphi.
\end{equation}
For $\delta>0$, the energy minimum is obtained with $\sigma_+$, and at 
$\cos^2 \varphi= (\pi \delta/2)^2 / \left[ \kappa^2 - \kappa (\pi \delta/2)^2 \right]$.
This value reaches 1 for
\begin{equation}
\delta= \delta_c \equiv \frac{2}{\pi} \frac{\kappa}{\sqrt{1+\kappa}}.
\end{equation}
This relation is a more general expression of the critical value of the DMI at which the
uniform state becomes unstable, as it takes better into account the internal magnetostatic energy
of N{\'{e}}el walls.
The expression at leading order is $\delta_c= 2 \kappa / \pi$ \cite{Thiaville12}; the
practical difference is weak as $\kappa$ vanishes in the limit of zero thickness
(for the samples studied here one indeed has $\kappa \simeq 0.1$).

\begin{figure}
\includegraphics[width= 8 cm]{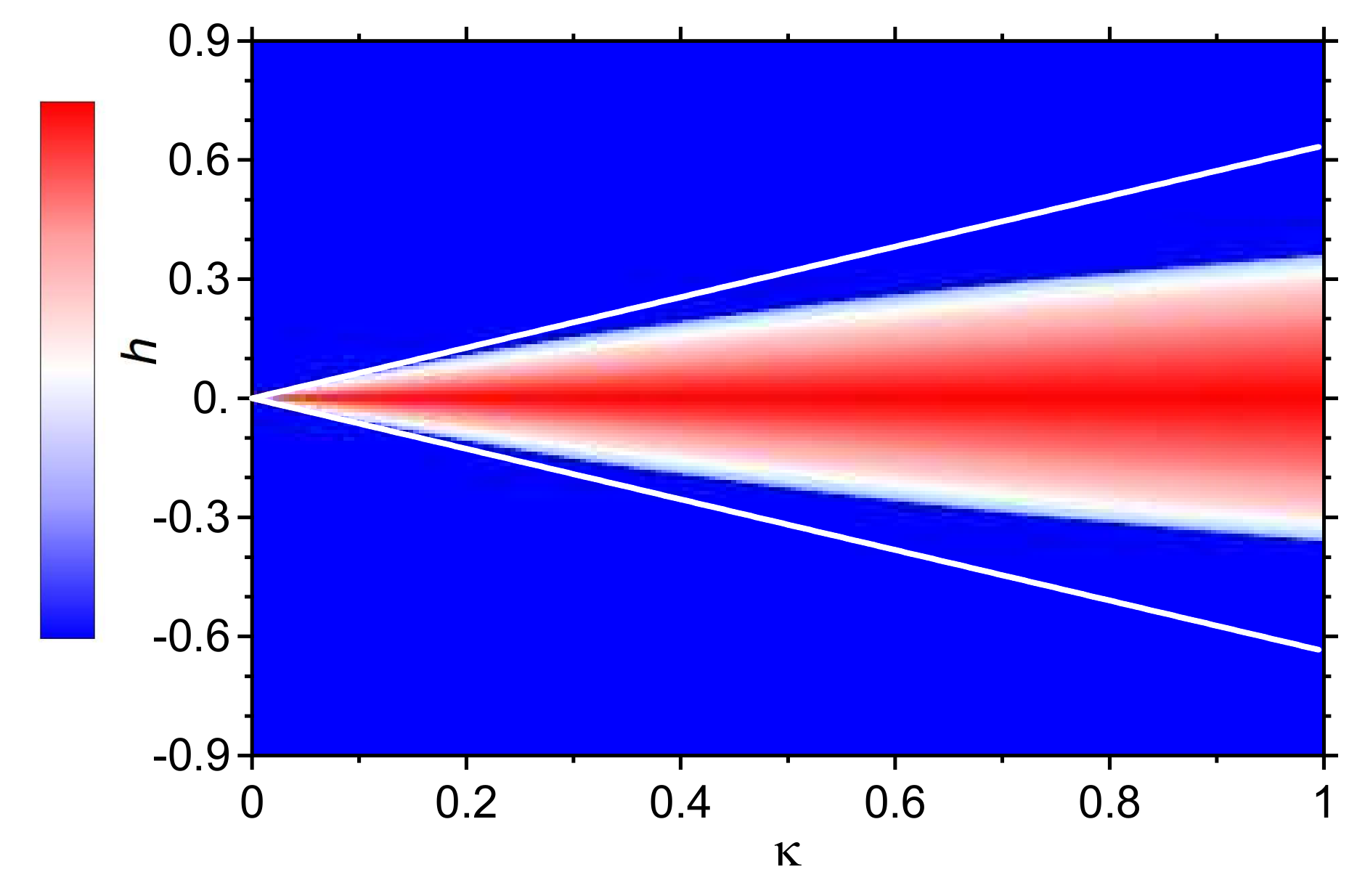}
\caption{
Results of the numerical minimization using the `small circle' Ansatz, for the
case of zero DMI, $\delta=0$.
The color code of the map shows the small circle cut angle $\varphi$,
it extends from 0 (blue) to $\pi/2$ (red).
The two lines (also obtained in the constrained model) depict the linear relation valid at small 
$\kappa$, see text.
\label{fig:carte-De0}
}
\end{figure}

Another limit is the a priori simple case with no DMI $\delta=0$.
Inspection of Eqns.~\ref{subeq:sig+} - \ref{subeq:sig-} shows that even in that case the minimization over
$\varphi$ is not simple.
The numerical solution of the problem is depicted in Fig.~\ref{fig:carte-De0}, as a map of the cut angle
$\varphi$ in the $(h, \kappa)$ plane.
For small domain wall anisotropy $\kappa$, the solution agrees with the simple expectation in the case where
the rotation of magnetization in the domains, and the deformation of the profile of the polar angle of the
magnetization across the domain wall, are neglected, namely $h= \pm (2/\pi) \kappa$ for the field required to
reach the N{\'{e}}el wall structure.

This simple case stresses that, even if the small circle model is easy to write down, its full solution
is not.
Therefore here stops, in general, the analytical work; one has to continue by a numerical minimization with 
respect to the cut angle $\varphi$.
It should be noted that, if the minimum is at $\varphi=0$, then the solution is exact (within the
assumption made for evaluating the energies).
The corresponding expressions for the domain wall energies were already given in \cite{Pizzini14,Vanatka15}.

\begin{figure}
\includegraphics[width= 8 cm]{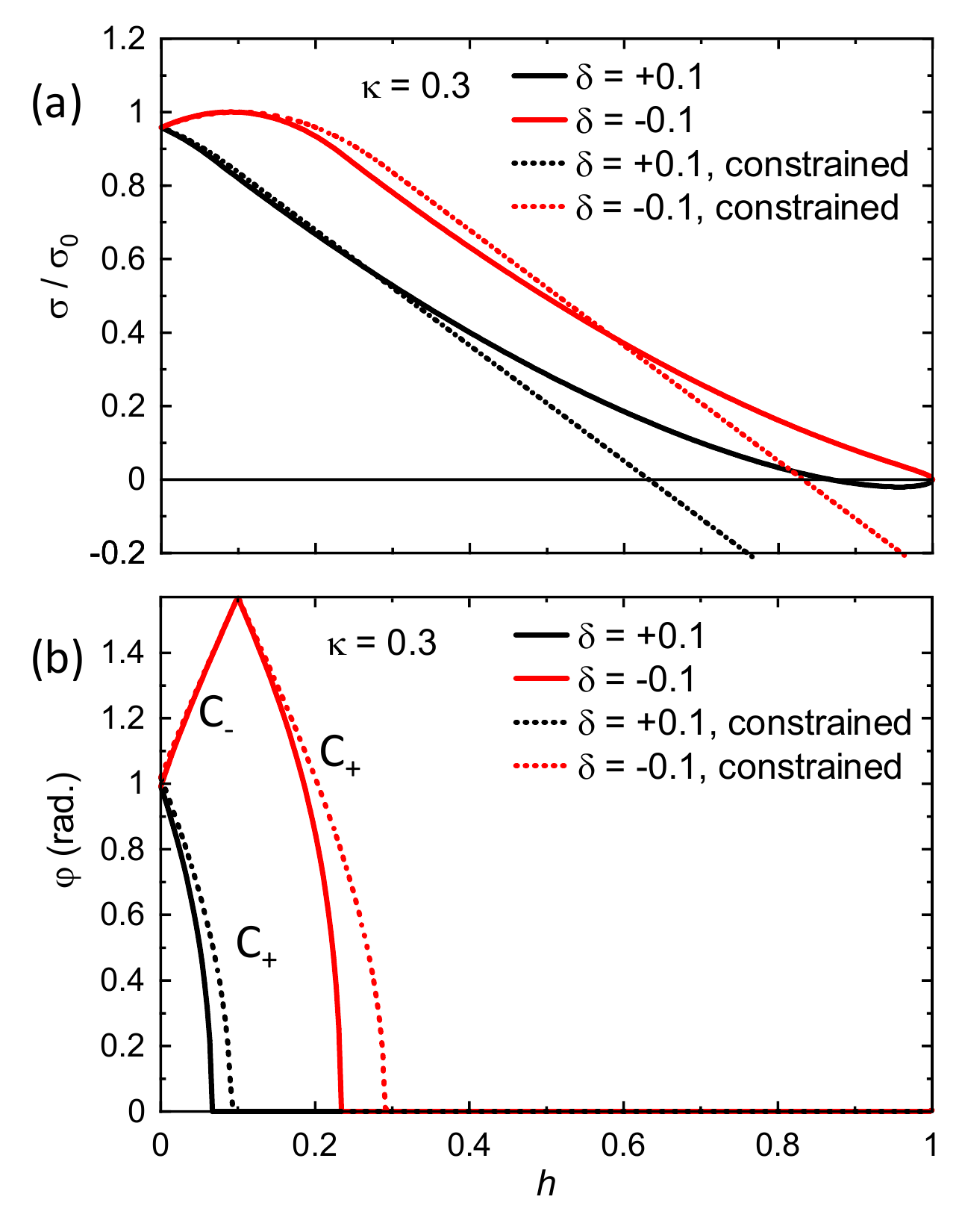}
\caption{
Results of the numerical minimization using the `small circle' Ansatz, for a large
domain wall relative magnetostatic energy ($\kappa=0.3$), for two opposite
and small values of the DMI ($\delta= \pm 0.1$).
(a) Normalized domain wall energy.
The energy maximum in the antiparallel case occurs at $h\approx 0.0901$.
(b) Corresponding cut angle $\varphi$, and minimum energy contours $C_\pm$
(see Fig.~\ref{fig:pc}).
The value $\varphi = \pi/2$ is reached at $h \approx 0.0995$.
The zero-field value $\varphi \approx 1$~radian means that the domain wall is in a mixed
Bloch-N{\'{e}}el state with the chosen values of $\kappa$ and $\delta$.
The results of the constrained model are drawn by dotted curves.
\label{fig:h-normal}
}
\end{figure}

To illustrate the model outputs, Fig.~\ref{fig:h-normal} shows the results for the case $\kappa=0.3$
and $\delta= \pm 0.1$, with the field acting in the same sense as DMI when $\delta > 0$,
whereas at zero field the wall is in an intermediate Bloch-N{\'{e}}el state.
The domain wall energies [Fig.~\ref{fig:h-normal}(a)] mostly decrease with field, and reach
0 at the effective anisotropy field where the domain wall vanishes as the magnetization turns in-plane,
parallel to the field.
In the case where DMI and applied fields are parallel, the characteristic negative domain wall
energy region \cite{Pizzini14} is obtained.
In the anti-parallel case, the domain wall energy reaches a maximum, before decreasing to 0.
This maximum is not exactly located at $h=-\delta$ (i.e. $H=-H_\mathrm{DMI}$), because the
domain wall magnetostatic energy is not negligible.

\begin{figure}
\includegraphics[width= 9 cm]{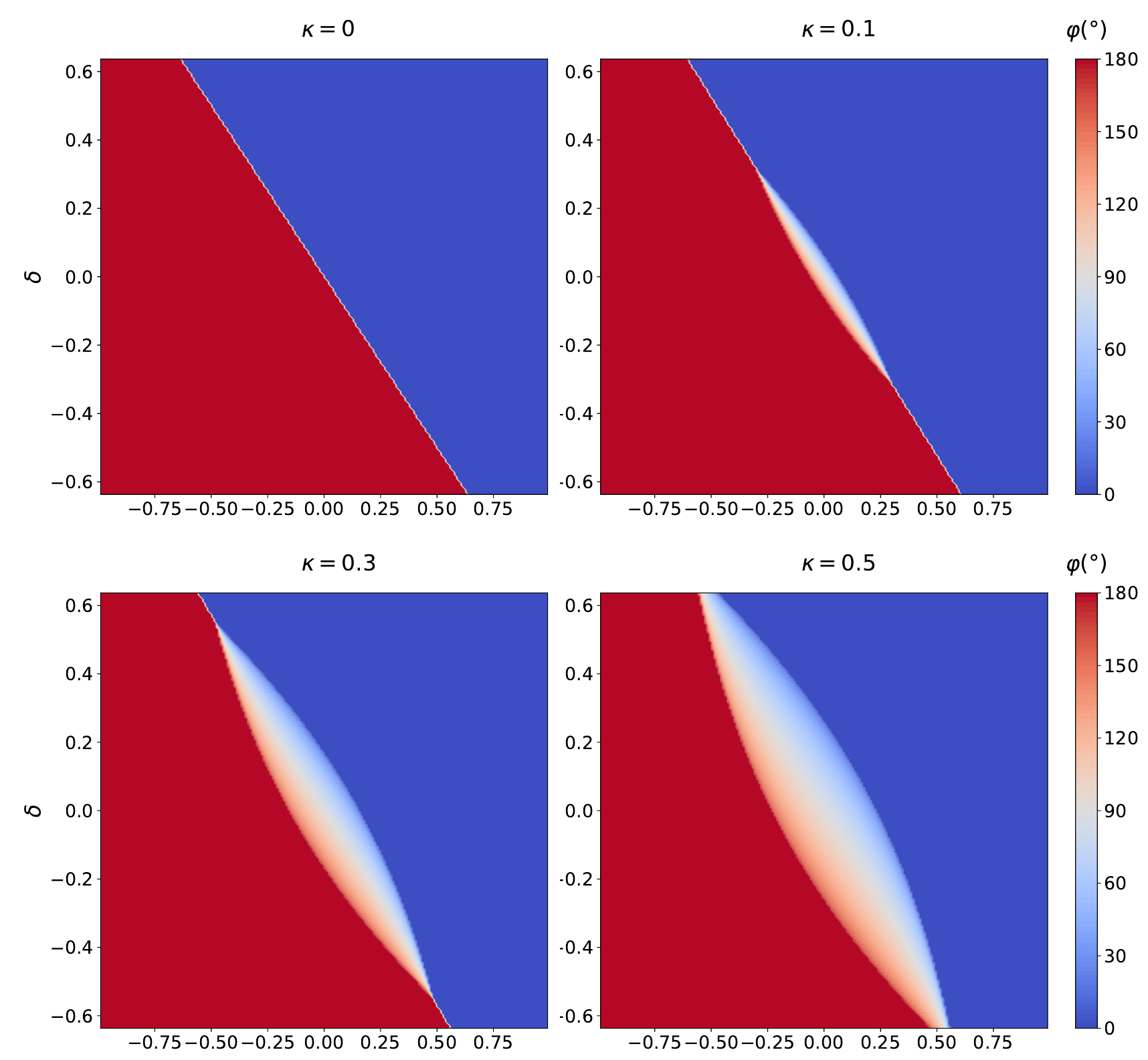}
\caption{
Type of domain wall as a function of in-plane field (scaled field $h$) and DMI (expressed by 
the scaled parameter $\delta$), for increasing values of the domain wall
magnetostatic energy (scaled parameter $\kappa$).
For these drawings, a fixed direction in the plane is considered, so that negative fields are 
figured, and $\varphi$ spreads over the $\left[0, \pi \right]$ interval.
The color code represents the cut angle $\varphi$, plotted over the extended range from 0
to $\pi$ so as to differentiate the two chiralities of N{\'{e}}el walls.
The constrained model predicts a switching at $h=-\delta$, the $\phi$ isovalues being parallels
to that line.
\label{fig:diagrammes}
}
\end{figure}

Repeating this calculations for various values of $\kappa$, $\delta$ and $h$, phase
diagrams can be constructed, as shown in Fig.~\ref{fig:diagrammes}.
They illustrate that the switching of the domain wall from one polarity of N{\'{e}}el
wall to the other takes place around $h=-\delta / \sqrt{1+\kappa}$, with a mixed
Bloch-N{\'{e}}el region that gets larger as $\kappa$ increases.
The above switching field relation can be obtained by equating $\sigma_+$ (Eq.~\ref{subeq:sig+})
to $\sigma_-$ (Eq.~\ref{subeq:sig-}) under the assumption that $\varphi=0$ (i.e. N{\'{e}}el walls).
The graphs show two switching processes, either continuous and through the Bloch wall in the vicinity
of the center of the graphs, or discontinuous from one N{\'{e}}el wall to the opposite one far
from the center of the graphs. 
One can obtain analytically the endpoints of the continuous region as
$(h, \delta)= \pm \left(-\sqrt{\kappa / (1+\kappa)} , \sqrt{\kappa} \right)$.

Once the domain wall profile is known, it is possible to compute some quantities of interest.
The first one is the so-called Thiele domain wall width $\Delta_\mathrm{T}$ which governs the domain
wall dynamics \cite{Thiele74,Thiaville06}.
It is defined by $2/\Delta_\mathrm{T} = \int{\left( d\vec{m}/dx \right)^2 dx}$.
One obtains
\begin{subequations}
\label{eq:lesDT}
\begin{eqnarray}
\frac{\Delta_\mathrm{T}^+}{\Delta_0} &=& \frac{1}{r^2 \sqrt{1+\kappa \cos^2 \varphi}
\left[ \cos\theta_0 -(\frac{\pi}{2}-\theta_0) \sin\theta_0 \right]} \label{subeq:DT+}\\
\frac{\Delta_\mathrm{T}^-}{\Delta_0} &=& \frac{1}{r^2 \sqrt{1+\kappa \cos^2 \varphi}
\left[ \cos\theta_0 +(\frac{\pi}{2}+\theta_0) \sin\theta_0 \right]} \label{subeq:DT-}
\end{eqnarray}
\end{subequations}
Note that, in the presence of an in-plane field that tilts the magnetization in the domains,
the famous steady-state velocity to easy-axis field ($H_z$) relation becomes
\begin{equation}
\label{eq:v-H}
v_x = \frac{\gamma_0 \Delta_\mathrm{T}}{\alpha} H_z \sqrt{1-h^2},
\end{equation}
(with $\gamma_0 \equiv \mu_0 |\gamma|$ the gyromagnetic factor and $\alpha$ the Gilbert
damping parameter, not to be confused with the angle of domain wall normal with field
resp. the domain wall tension, both used in Sec.~\ref{sec:gene}).
Thus, the velocity increase due to that of the Thiele DW width is partly compensated by the decrease
of the driving force due to the domain magnetization tilt, illustrating the fact that the in-plane 
field has conflicting influences on the domain wall mobility.

The other domain wall width of interest is the `imaging' width, introduced by
A.~Hubert \cite{Hubert74,Jue16}, which measures the extension in physical space
of the domain wall.
It is anticipated that this width is the relevant one to evaluate the pinning of 
the domain wall by imperfections \cite{Gehanne20}.
Using the general definition of Ref.~\cite{Jue16}, based on the value of
$m_z$ at infinity, one gets
\begin{subequations}
\label{eq:lesDH}
\begin{eqnarray}
\frac{\Delta_\mathrm{H}^+}{\Delta_0} &=& \frac{\cos\theta_0}
{\sqrt{1+\kappa \cos^2 \varphi} \left[1 - \sin\theta_0 \right]} \label{subeq:DH+}\\
\frac{\Delta_\mathrm{H}^-}{\Delta_0} &=& \frac{\cos\theta_0}
{\sqrt{1+\kappa \cos^2 \varphi} \left[1 + \sin\theta_0 \right]} \label{subeq:DH-}
\end{eqnarray}
\end{subequations}
\begin{figure}
\includegraphics[width= 8 cm]{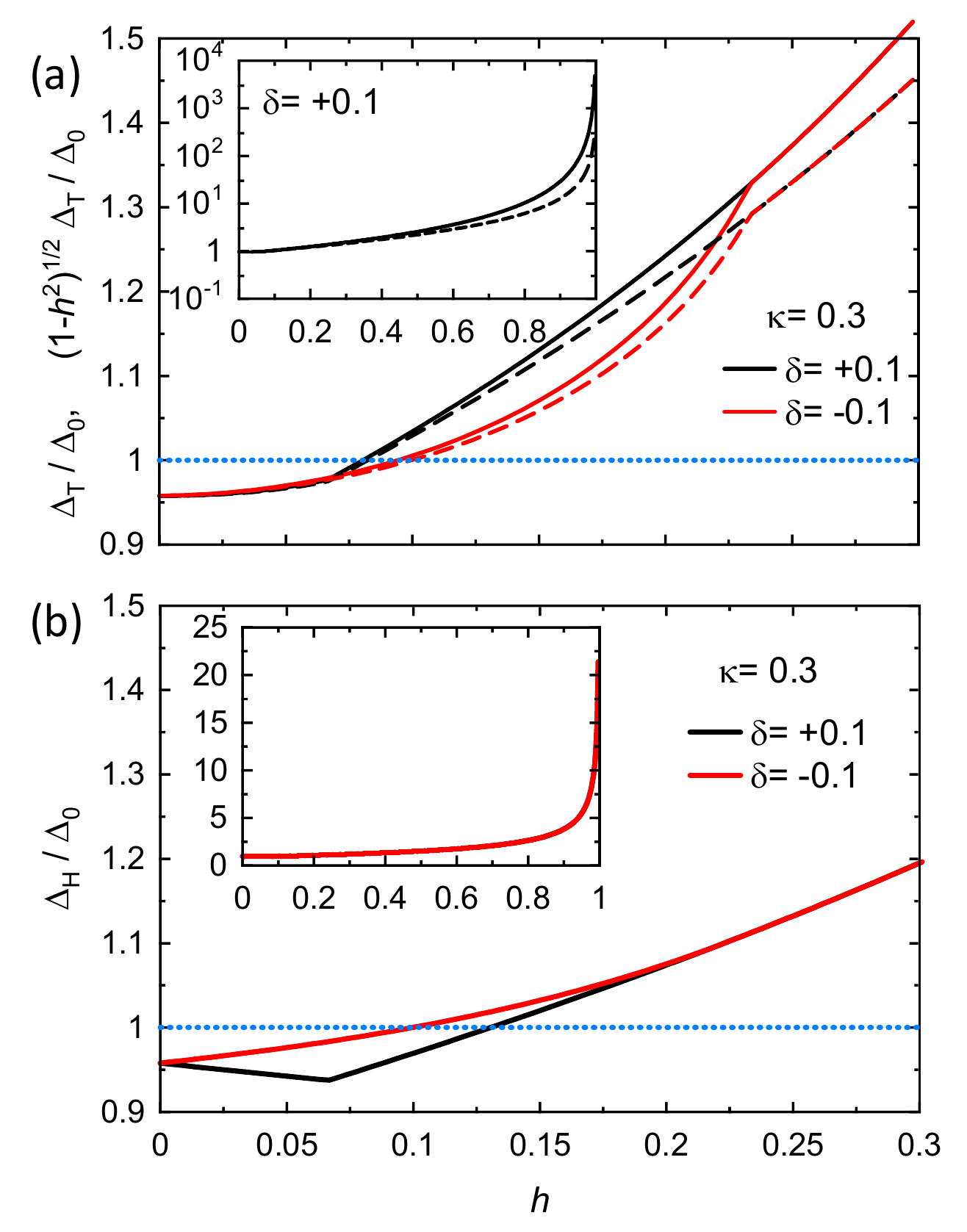}
\caption{
Results of the numerical minimization using the `small circle' Ansatz, for a large
domain wall relative magnetostatic energy ($\kappa=0.3$), for two opposite
and small values of the DMI ($\delta= \pm 0.1$).
(a) Thiele domain wall width $\Delta_\mathrm{T}$ normalized to $\Delta_0$, for the small fields.
The inset shows the full curve, in log-lin scale.
The dashed curves plot the product 
$\sqrt{1-h^2} \Delta_\mathrm{T}$, the relevant quantity for the domain wall mobility.
(b) Hubert domain wall width $\Delta_\mathrm{H}$, normalized to $\Delta_0$, the inset
showing the full curve, this time in lin-lin scale.
The breaks of the curves correspond to the fields ($h=0.0664$ for $\delta=+0.1$
and $h=0.2334$ for $\delta=-0.1$) where the domain wall
becomes of the N{\'{e}}el type, perfectly parallel to the field [$\varphi= 0$,
see Fig.~\ref{fig:h-normal}(b)].
The constrained model assumes that all widths are constant, equal to $\Delta_0$ (dotted lines).
\label{fig:deltas}
}
\end{figure}

These two widths are plotted in Fig.~\ref{fig:deltas}; they globally increase
with the in-plane field, as expected.
The width smaller than $\Delta_0$ at zero field expresses the contraction due to
the magnetostatic cost of the non fully Bloch wall.
Whereas the Thiele domain wall width diverges at $H= H_{K0}$ (as no
magnetization gradient anymore exists at that field), the Hubert width increases less.
In the intermediate field region (intermediate Bloch- state), the width
differs for fields parallel and antiparallel to the DMI field.
Moreover, in the case of large domain wall anisotropy and low DMI, the Thiele
and Hubert widths can show opposite trends with field [Fig.~\ref{fig:deltas}(b)].
This is due to the behavior of the various factors entering the domain wall
widths, see Eqs.~(\ref{eq:lesDT},\ref{eq:lesDH}).
Especially, the DW width can decrease with field when the factor containing the 
DW anisotropy (parameter $\kappa$) is dominant, the DW magnetization turning
from Bloch to N{\'{e}}el as field increases.

Note also that, for this relatively small DMI compared to the domain wall internal
magnetostatic energy, no minimum of the DW width occurs at $H_x= - H_\mathrm{DMI}$,
in contrast with what is predicted by simplified models.
Thus, the small circle Ansatz allows an exploration of the complex physics of
the statics of domain walls submitted to a transverse field, in which
several effects are in competition.

\subsection{Comparison to numerical micromagnetics}
\label{sec:mumag-normal}

\begin{table}[t]
\caption{Parameters of the three samples investigated, all with a
nominal 0.9~nm Co thickness (the name reflects the growth order).
The exchange constant is assumed to be $A= 16$~pJ/m.}
\label{tab:params}
\begin{tabular}{ c c | c c c }
\hline \hline
Sample & & Au/Co/Pt & Pt/Co/Au & Pt/Co/Pt \\
\hline
$M_\mathrm{s}$  & (kA/m) & 1610   &  1650   &  1621     \\
$K_u$ & (MJ/m$^3$) & 2.36 & 2.35 & 2.12 \\
$D$ & (mJ/m$^2$) & 0.60 & -0.87 & 0 \\
\hline
$\mu_0 H_{K0}$ & (T) & 0.9 & 0.77 & 0.58 \\
$\Delta_0$ & (nm) & 4.7 & 5.0 & 5.8   \\
$\kappa$ & () & 0.10  & 0.11 &  0.12  \\
$\delta$ & () & 0.09   & -0.20 & 0 \\
\hline \hline
\end{tabular}
\end{table}

We now compare quantitatively the results of the model to those obtained by
numerical micromagnetic calculations, using MuMax3 \cite{Vansteenkiste14}.
For these calculations, the sample was meshed in $1024 \times 1 \times 1$ cells in the $x$,
$y$ resp. $z$ directions, with a cell size $1\times 1 \times 0.9$~nm$^3$
and periodic boundary conditions in the $y$ direction, the magnetostatic interaction
coefficients being summed over 100 000 repetitions (this value was reached by
comparison with the analytical demagnetizing factor, for a uniform magnetization).
To avoid edge effects, known to exist with DMI \cite{Rohart13}, the data for the
domain wall were collected on the 400 central cells, this length being also well
above the obtained domain wall widths.
The magnetic parameters of the three samples considered in these calculations are
provided in Tab.~\ref{tab:params} (see ref.~\cite{Gehanne20} for details).
Important parameters derived from these values are also given, in particular the
numbers $\kappa$ and $\delta$.

\begin{figure}
\includegraphics[width= 8 cm]{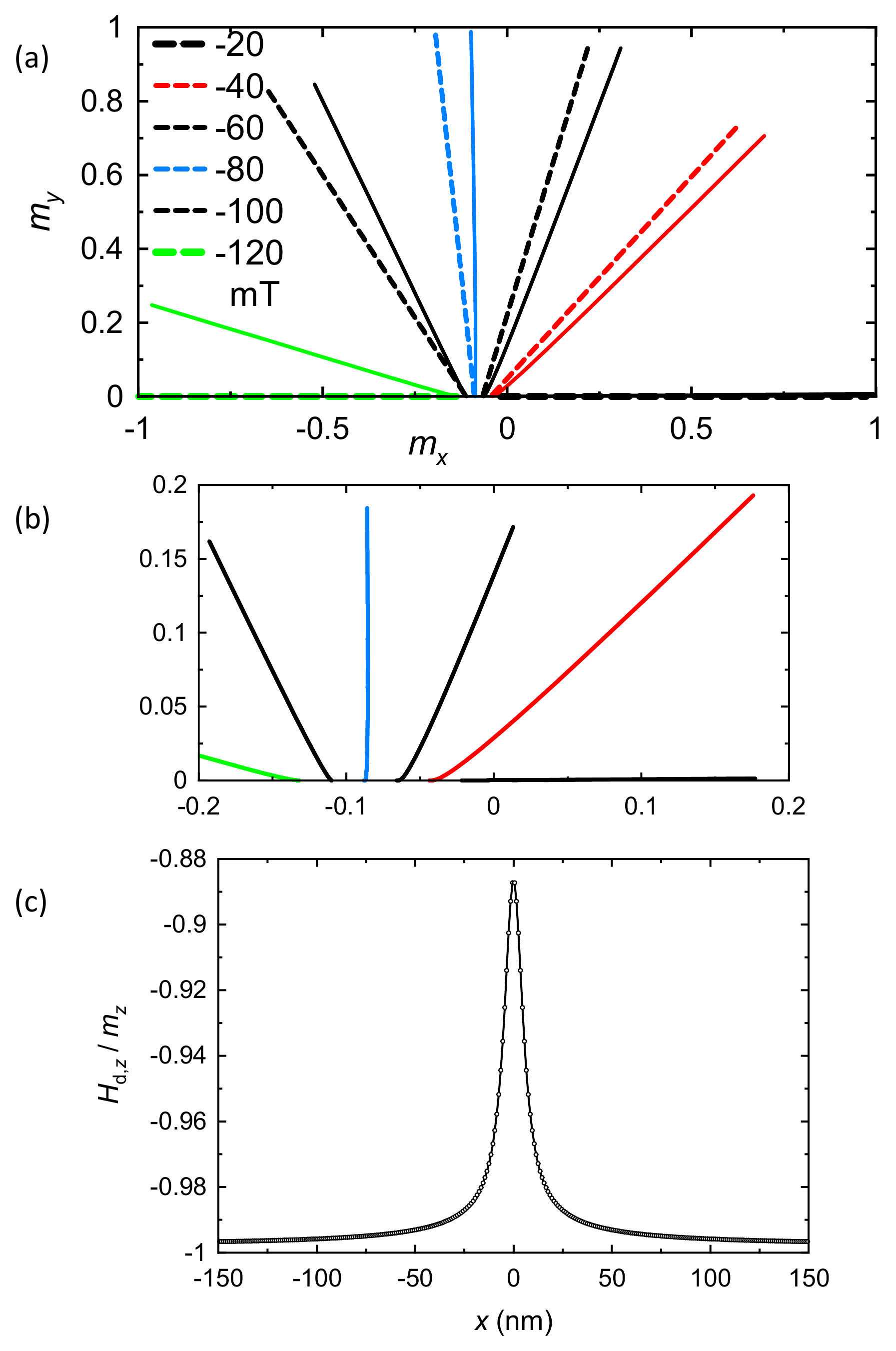}
\caption{
Comparison of the small circle model with numerical simulations, for the
Au/Co/Pt sample.
(a) The in-plane magnetization components for an up-down domain wall
under several in-plane fields, predicted by the small circle model 
(dashed lines) and computed by numerical micromagnetics (curves).
(b) Zoom close to the origin, for the numerical profiles, to
evidence the small deviations from linearity.
(c) Profile of the local $z$ demagnetizing factor as extracted from
the micromagnetic calculation under zero applied field.
\label{fig:ACP-prof}
}
\end{figure}

The profiles of the domain wall magnetization projected on the $(m_x, m_y)$ plane
(this plane was used to draw Fig.~\ref{fig:pc}a)
are shown in Fig.~\ref{fig:ACP-prof} (a), as obtained by the small circle model (dashes),
and by numerical micromagnetics (continuous curves), for several values of the in-plane field that
span reversal of the domain wall magnetic moment.
One first notes that the small circle approximation is very good, as the numerical
profiles are extremely close to straight lines, the traces on the $(m_x,m_y)$ plane
of the vertical cut plane.
Small deviations to this behavior are seen close to the origin of the plots, i.e. at the
tails of the domain wall [see Fig.~\ref{fig:ACP-prof}(b)].
However, the cut angles are found to differ.
This is due to the model's assumption of a purely local $z$-component of the demagnetizing field, 
namely $H_{d,z}=-M_\mathrm{s} m_z$. 
Plot (c) shows that indeed the demagnetizing field perpendicular component falls below the $-1$ ratio
to the normal magnetization component $m_z$, at the domain wall.
As a result, the Thiele domain wall width in zero applied field, estimated to
be 4.7~nm in the model, is 4.37~nm in the numerical calculation.
This leads to a larger demagnetizing cost of the N{\'{e}}el wall, so that
the reversal of the domain wall $x$ magnetic moment extends over a larger field region.

\begin{figure}[t]
\includegraphics[width= 8 cm]{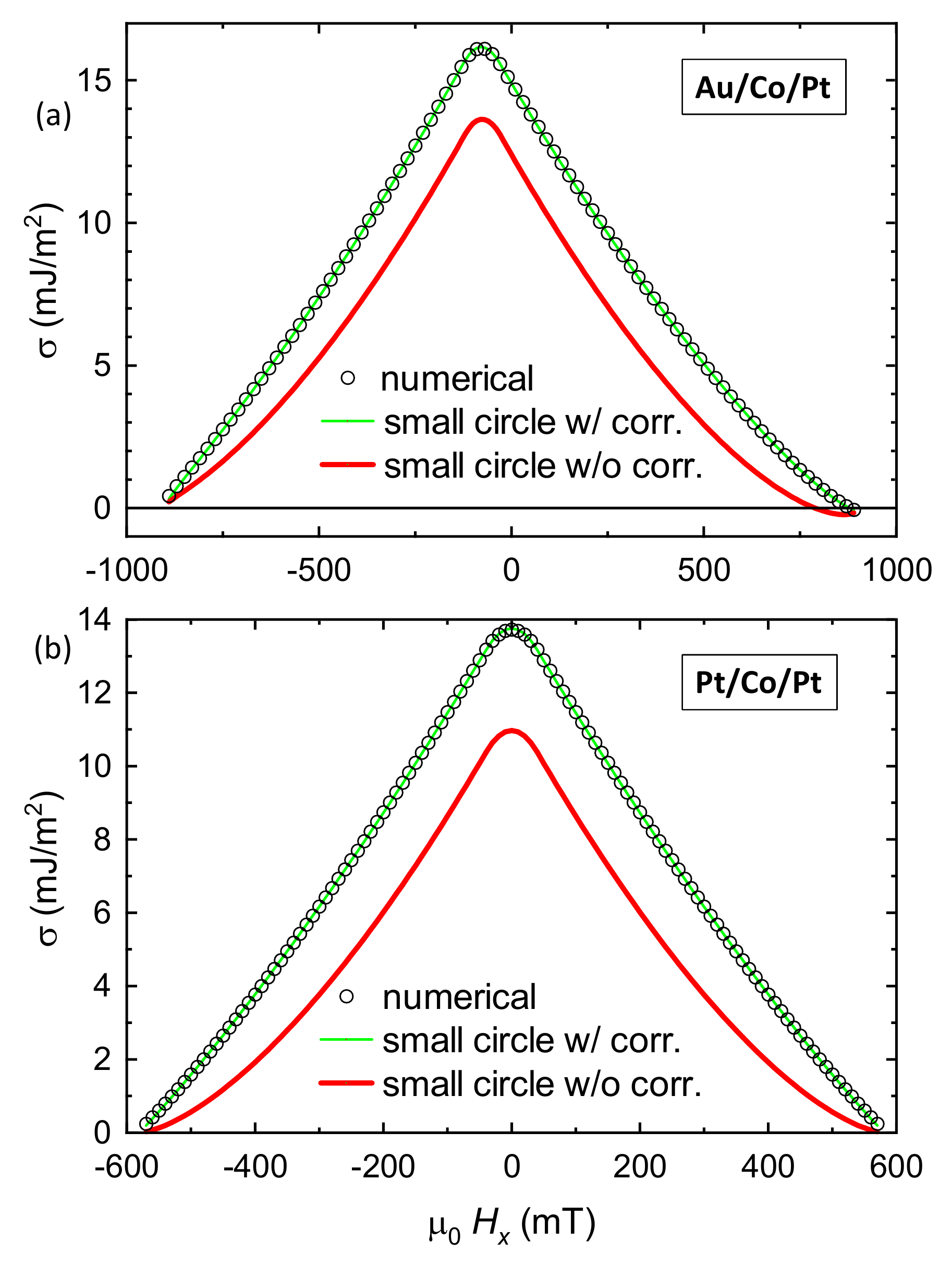}
\caption{
Comparison of the small circle model with numerical simulations, for
the Au/Co/Pt (a) and Pt/Co/Pt (b) samples, regarding the domain wall energy.
The effect of the analytical domain wall energy correction (see Appendix)
is shown.
For Au/Co/Pt, the domain wall energy maximum is reached at $\mu_0 H_x=-75$~mT.
\label{fig:compa-sigmas}
}
\end{figure}

Next we look at the energy of the domain wall, per unit surface \cite{note-sigma}.
As visible in Fig.~\ref{fig:compa-sigmas}, the small circle model gives
precisely the same trend as the numerical simulation, but lower values
due to the neglect of the non-local magnetostatic term within the domain wall.
A calculation detailed in the Appendix leads to a correction term, dependent on the width
of the domain wall, which corrects most of this difference (see Fig.~\ref{fig:compa-sigmas}).

\begin{figure}
\includegraphics[width= 8 cm]{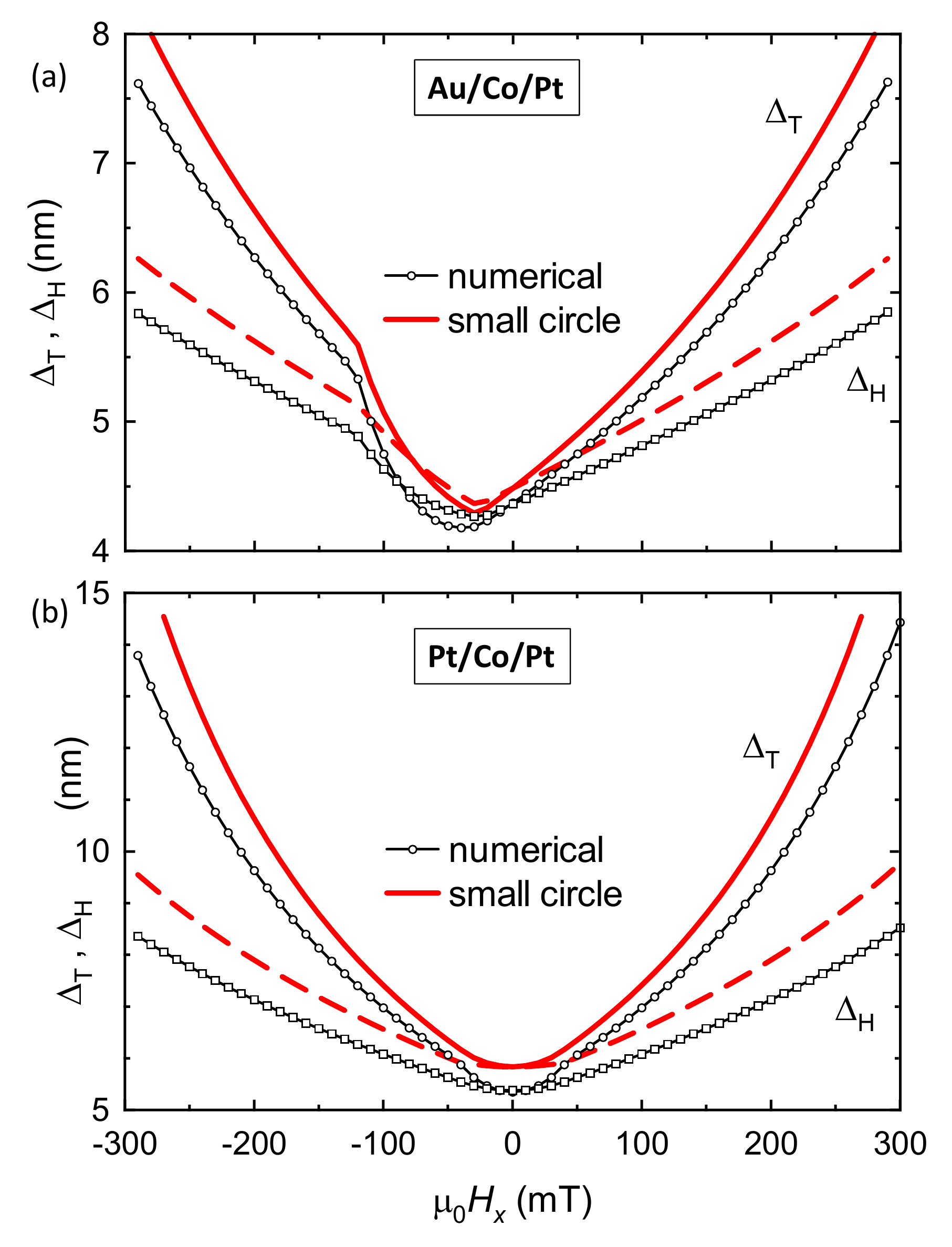}
\caption{
Comparison of the small circle model with numerical simulations, for
the AuCoPt (a) and PtCoPt (b) samples, regarding the domain wall widths
(see text for definitions).
For AuCoPt, all widths show a minimum at -30~mT for the small circle model, and
-40~mT for the numerical calculation, values which are clearly
lower from that where the domain wall energy is maximum 
[see Fig.~\ref{fig:compa-sigmas}(a)].
\label{fig:compa-Deltas}
}
\end{figure}

Finally the widths of the domain wall, either relevant for dynamics (the
Thiele domain wall width parameter $\Delta_\mathrm{T}$) or for imaging
(the Hubert domain wall width parameter $\Delta_\mathrm{H}$) are investigated.
The comparison of the numerical micromagnetic simulation results with those of the small
circle model is detailed in Fig.~\ref{fig:compa-Deltas}.
One sees that, generally, the widths are larger with the small circle model, due
to the local approximation for the $z$ component of the demagnetizing field, as well
as for the $x$ component [Eq.~(\ref{subeq:eBN})]. 
The computed variations of the domain wall width are important in relative terms.
Even if it changes, in absolute terms, only between 4 and 6~nm, an effect on the domain wall pinning
characteristics has been experimentally observed \cite{Gehanne20}.

One notices that the widths, in this case where DMI is comparable to or even larger than
the domain wall internal magnetostatic energy, show a minimum for some field opposite
to the DMI field.
This field is however smaller, in absolute value, than the field where the domain wall energy 
is maximum.
Therefore, the various methods based on domain walls to measure the interfacial DMI should be 
compared in detail.

Globally, the small circle model is shown to be quite accurate, for the magnetization
profile, its spatial extent and its energy.
The biggest difference appears to lie in the domain wall energy.
Most of it may be corrected by adding an estimate of the non-local magnetostatic
energy of the domain wall.
%


\section{Field at an arbitrary angle}
\label{sec:gene}

For this general case, we proceed similarly to the previous part.
Only what changes or was not present in the high symmetry case is given.

\begin{figure*}
\includegraphics[width= 15 cm]{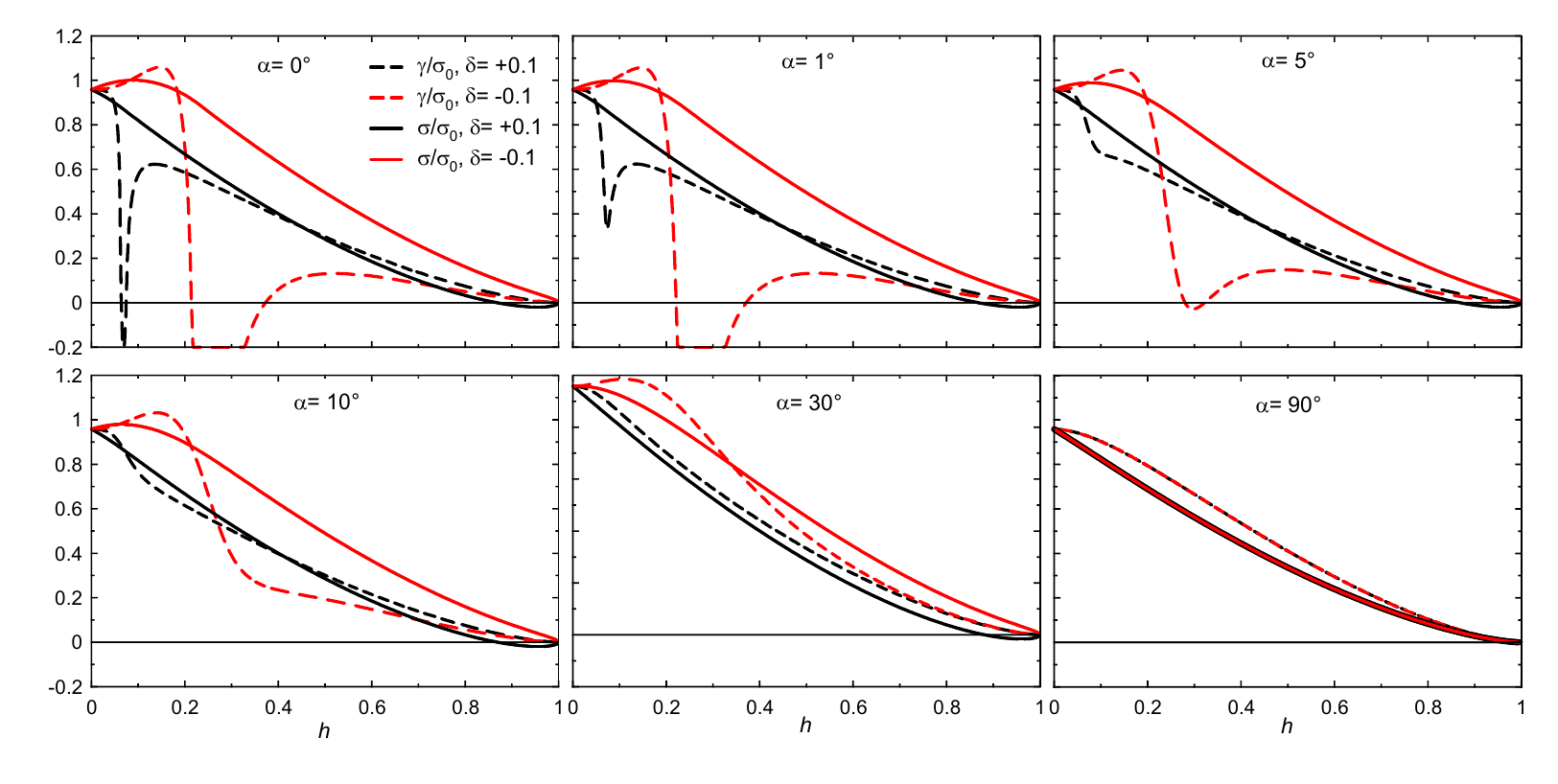}
\caption{
Results of the numerical minimization using the `small circle' Ansatz, for a large
domain wall relative magnetostatic energy ($\kappa=0.3$), for two opposite
and small values of the DMI ($\delta= \pm 0.1$).
The domain wall surface energy $\sigma$ (lines) as well as surface tension $\gamma$ (dash lines),
 normalized to the Bloch wall surface energy $\sigma_0$, are plotted as a function of the normalized
in-plane field $h$, for several values of the angle $\alpha$ between the domain wall normal
and the applied field.
\label{fig:sig&gam-h}
}
\end{figure*}
\begin{figure*}
\includegraphics[width= 15 cm]{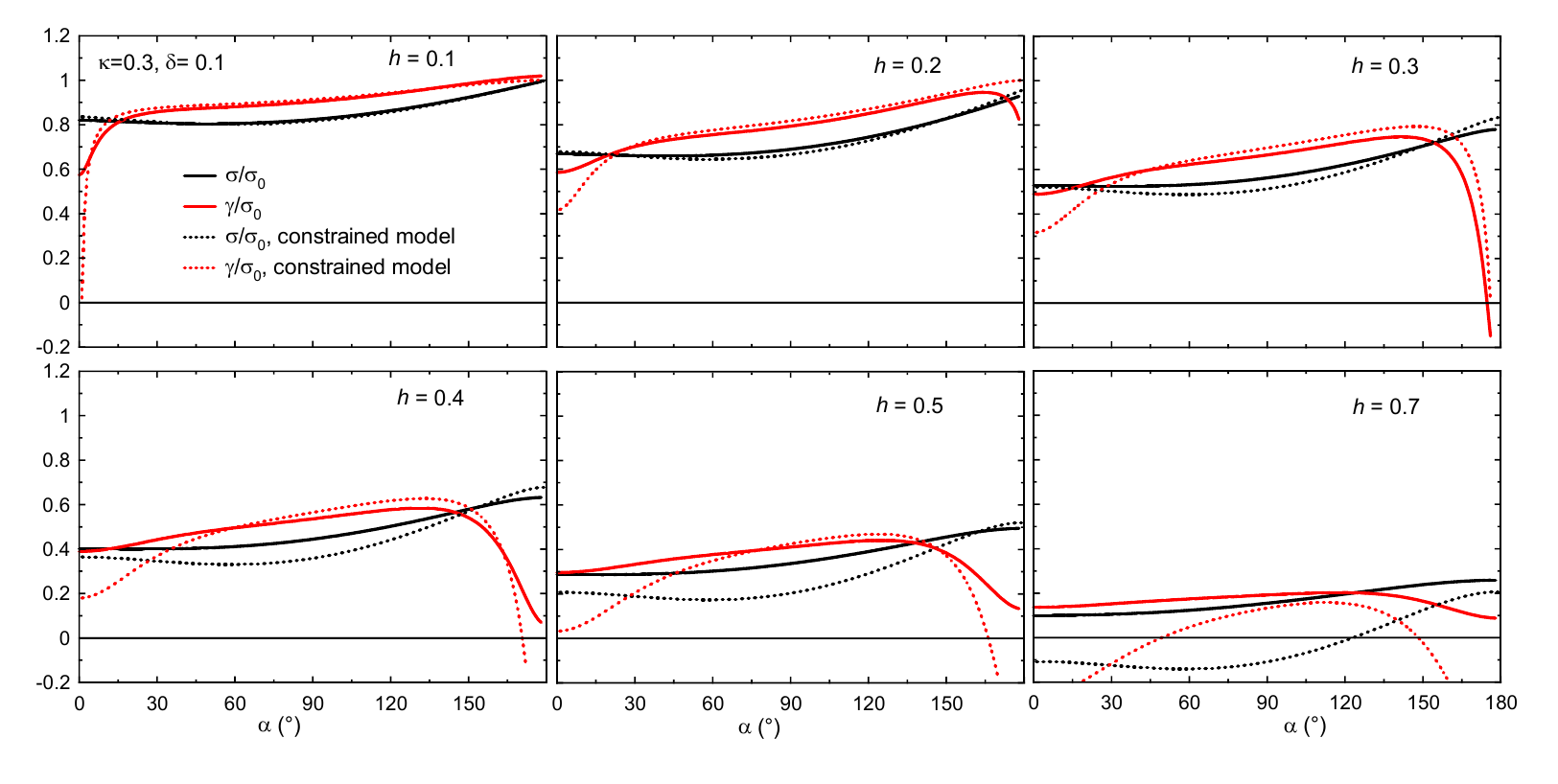}
\caption{
Alternative view of the results shown in Fig.~\ref{fig:sig&gam-h}, for $\kappa= +0.3$ and
$\delta= +0.1$, where the angle $\alpha$ of the domain wall normal with the in-plane field is varied, 
for selected values of the field.
The results of the constrained model have been included (dotted curves) for comparison.
\label{fig:sig&gam-a}
}
\end{figure*}

\subsection{Semi-analytical model}
\label{subsec:ana-angle}

The angle of the domain wall normal $\vec{n}$ with the field ($x$) axis is called $\alpha$.
The energy densities which depend on the domain wall orientation are the DMI and domain wall
internal magnetostatic energies. 
They read
\begin{eqnarray}
\label{eq:lesE2}
&\mathcal{E}_\mathrm{DMI}& = -D \frac{d\theta}{dn} r \left[ h \sin \varphi \sin(\varphi - \alpha) \sin\theta 
+ r \cos(\varphi-\alpha) \right]  \nonumber \\
&\mathcal{E}_\mathrm{BN}& = K \cos^2 (\varphi-\alpha) \left( h \cos\varphi - r \sin\theta \right)^2.
\end{eqnarray}
The same analysis as before leads to the variation law of the angle $\theta$
\begin{equation}
\label{eq:dtdx2}
\frac{d \theta}{d \xi}= \sqrt{1+\kappa \cos^2 (\varphi-\alpha)} \left( \sin\theta - \sin\theta_0 \right).
\end{equation}
Similarly, the domain wall energies for both arcs are
\begin{subequations}
\label{eq:les-sig2}
\begin{eqnarray}
&&\frac{\sigma_+}{\sigma_0} = r^2 \sqrt{1+\kappa \cos^2 (\varphi-\alpha)} \left[ \cos\theta_0 - 
(\frac{\pi}{2}-\theta_0) \sin\theta_0 \right]  \nonumber \\
&&- r \delta \left[ h \sin\varphi \sin(\varphi-\alpha) \cos\theta_0
+(\frac{\pi}{2}-\theta_0)r \cos(\varphi-\alpha) \right], \nonumber \\
&& \label{subeq:sig2+} \\
&&\frac{\sigma_-}{\sigma_0} = r^2 \sqrt{1+\kappa \cos^2 (\varphi-\alpha)} \left[ \cos\theta_0 + 
(\frac{\pi}{2}+\theta_0) \sin\theta_0 \right] \nonumber \\
&&- r \delta \left[ h \sin\varphi \sin(\varphi-\alpha) \cos\theta_0
-(\frac{\pi}{2}+\theta_0)r \cos(\varphi-\alpha) \right]. \nonumber \\
&& \label{subeq:sig2-}
\end{eqnarray}
\end{subequations}
Finally, the domain wall widths have similar expressions to the normal case 
[Eqs.~(\ref{eq:lesDT}, \ref{eq:lesDH})], only replacing $\varphi$
by $\varphi - \alpha$ inside the square root with $\kappa$.
These formulas illustrate the power of the model: treating a much more general problem is
realized by a minor modification of the functions to use.
For the constrained model, counting the domain wall magnetization angle $\psi$ from the
domain wall normal, the function to minimize, for each value of $\alpha$, reads now
$\sigma(\psi)/\sigma_0=1+(\kappa/2) \cos^2\psi -(\pi/2)[\delta \cos\psi
+ h\cos(\psi-\alpha)]$.

The computed dependence of the domain wall energy on the domain wall orientation $\alpha$ allows
evaluating another important parameter, the domain wall surface tension $\gamma$.
It is defined by
\begin{equation}
\label{eq:gam}
\gamma (\alpha) = \sigma(\alpha) + \frac{d^2 \sigma}{d \alpha^2}.
\end{equation}
Indeed, in the case of a domain wall surface energy that depends on the domain wall orientation,
the energy cost of a bulging of the domain wall consists of (i) the increase of the 
wall length, penalized by $\sigma$, and (ii) an energy variation due to the exploration of neighboring
domain wall angles by the bulge, which leads to the second derivative.
The distinction between energy and tension is a well-known concept in surface physics
\cite{Desjonqueres96}, and its relevance for magnetic domain wall creep motion was recently 
stressed \cite{Pellegren17,Hartmann19}.

\begin{figure*}[t]
\includegraphics[width= 15 cm]{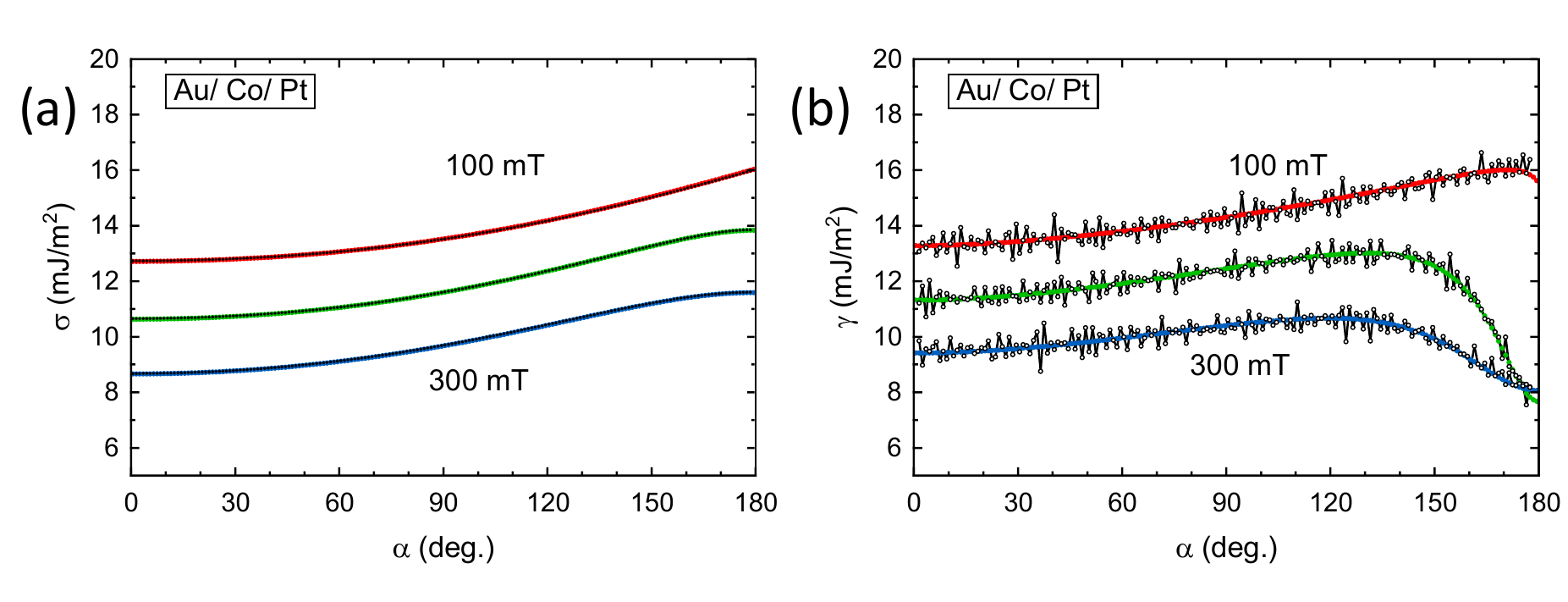}
\caption{Comparison of the numerical micromagnetics results (open black symbols) to those
of the semi-analytical small circle model (colored curves), for the Au/Co/Pt sample.
(a) variation of the domain wall energy $\sigma$ with angle of the applied in-plane field,
for 3 values of the field $\mu_0 H= 100, 200, 300$~mT.
(b) variation of the domain wall tension $\gamma$ in the same conditions.
The analytical correction to the domain wall energy (and tension) is included in the
small circle values.
Note the absence of noise in the semi-analytical calculation of the tension, compared
to the numerical procedure (see text for cause).
\label{fig:compa-gamma}
}
\end{figure*}
\begin{figure*}
\includegraphics[width= 15 cm]{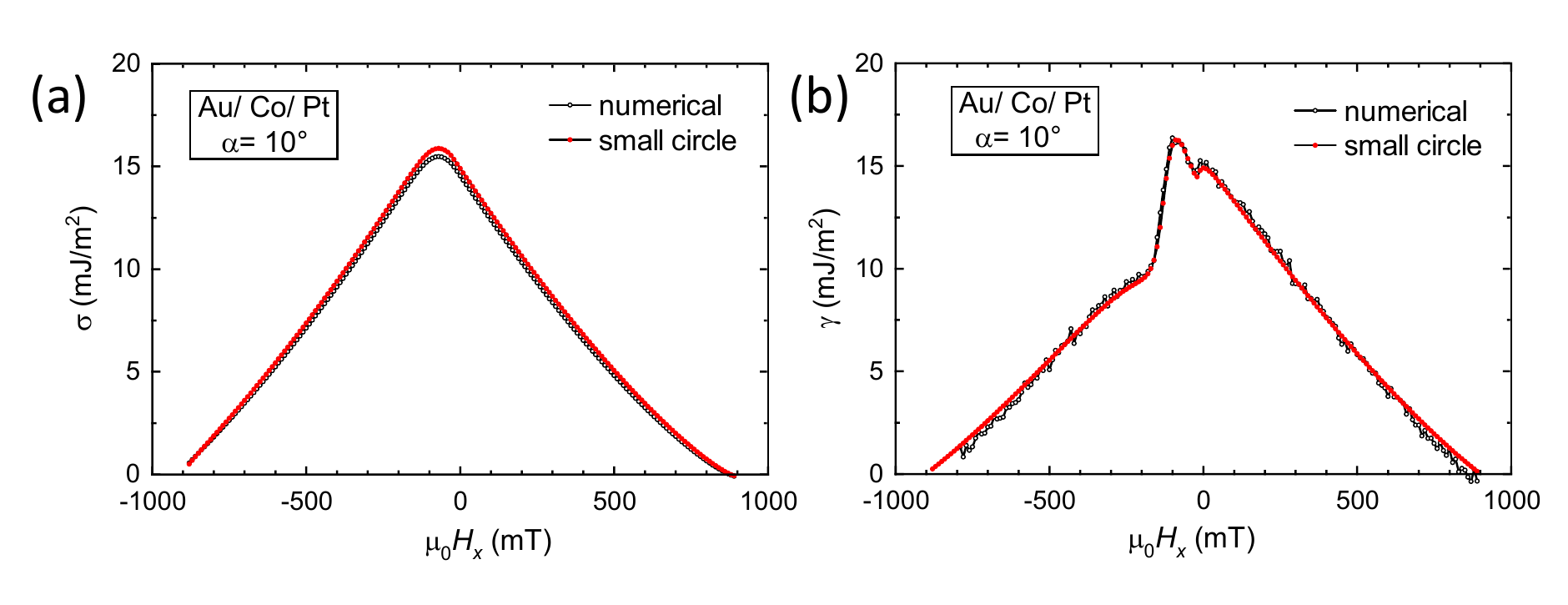}
\caption{Comparison of the numerical micromagnetics results (open black symbols) to those
of the semi-analytical small circle model (red curves), for the Au/Co/Pt sample.
(a) variation of the domain wall energy $\sigma$ with the in-plane field, applied at an angle
$\alpha= 10$ degrees.
(b) variation of the domain wall tension $\gamma$ in the same conditions.
The analytical correction to the domain wall energy (and tension) is included in the
small circle values.
\label{fig:compa-gamma-2}
}
\end{figure*}

For the demonstration sample with large domain wall internal magnetostatic energy ($\kappa=0.3$) 
first, the variation of domain wall surface energy and tension as a function of in-plane field and
domain wall orientation is shown in Fig.~\ref{fig:sig&gam-h}.
For large angles ($\alpha > 20^\circ$), domain wall energy and tension follow similar
evolutions with applied field.
One should nevertheless note that the negative domain wall energies found close to saturation
give positive tensions.
At low angles however, things are much more complex, with the appearance of regions in the
$\left( \alpha , h \right)$ space where the domain wall tension is negative.
This situation is well known in crystal growth \cite{Desjonqueres96}: such a domain
wall orientation is unstable, and faceting appears (a phenomenon also called the
zig-zag faceting in magnetism \cite{Hubert74}).
The faceting in the case where in-plane field is parallel to DMI field ($\delta > 0$ here)
is minute, as the domain wall tension is already postive at $\alpha= 1^\circ$ [for $h= 0.07$
for example, the tension reaches 0 at $\alpha \equiv \alpha_\mathrm{c}=0.515^\circ$ and the faceting 
occurs with angles ($+$ or $-$) $\alpha_\mathrm{f}= 1.29^\circ$].
It is larger in the antiparallel case: at $h=0.3$ the angles are $5.25^\circ$ and $\pm 11.3^\circ$,
respectively.
In the $\gamma <0$ region (i.e. $|\alpha| < \alpha_\mathrm{c}$), the domain wall energy
$\sigma$ is replaced by 
$\sigma_\mathrm{f}(\alpha)= \sigma(\alpha_\mathrm{f}) \cos\alpha / \cos\alpha_\mathrm{f}$,
with a discontinuity at $\alpha=\alpha_\mathrm{c}$.
This relation moreover leads to a domain wall tension which is exactly zero.
These conclusions hold, however, only in the infinite domain wall length limit, as the energy
cost of the kinks of the faceted domain wall is not taken into account.

Another noticeable feature is that, for moderate angles ($|\alpha| < 20^\circ$), at non-negligible
fields ($h > 0.25$) the favored domain wall has a lower energy, but a larger tension.

The other way to look at the same data, in which the angle $\alpha$ varies continuously, is 
shown in Fig.~\ref{fig:sig&gam-a}.
In addition, the results of the constrained model have been included in these graphs (dashed curves).
The differences with the small circle model steadily increase as the field becomes larger, and they are more 
important for the domain wall tension $\gamma$, with larger variations with angle predicted by the constrained model.
This should be expected, from the presence of the second derivative of the domain wall energy versus angle,
which is sensitive to the fine variations of the domain wall energy $\sigma$.
The large differences of computed domain wall tension mean large differences of the domain wall mobility
as a function of field orientation, hence for example big differences of shape of bubble domains when expanding
in the creep regime in the presence of an in-plane field \cite{Pellegren17}.

\subsection{Comparison with numerical micromagnetics}
\label{sec:mumag-angle}

We now turn to the samples investigated in this study.
Using the same procedure, the domain wall surface energy $\sigma$ was numerically evaluated. 
In order to obtain the domain wall tension $\gamma$, a finite differences evaluation of
the second derivative versus angle was performed.
Due to the limited precision of the numerical values (single precision), the angle step could
not be reduced below 1 degree.
For the small circle calculations, a much smaller angle step could be used ($10^{-5}$ degree), 
as the calculations are performed with double precision, resulting into a smaller numerical 
noise.
The analytical correction of the domain wall energy (see Appendix) was added to the small circle 
model results, using for the domain wall width parameter the Hubert value.
As the variation with angle of the Hubert domain wall width is small (for example, $\pm 0.1$~nm
around 5.7~nm for Au/Co/Pt at 200~mT), this correction amounts to the same offset for $\sigma$ 
and $\gamma$.

\begin{figure}
\includegraphics[width= 8 cm]{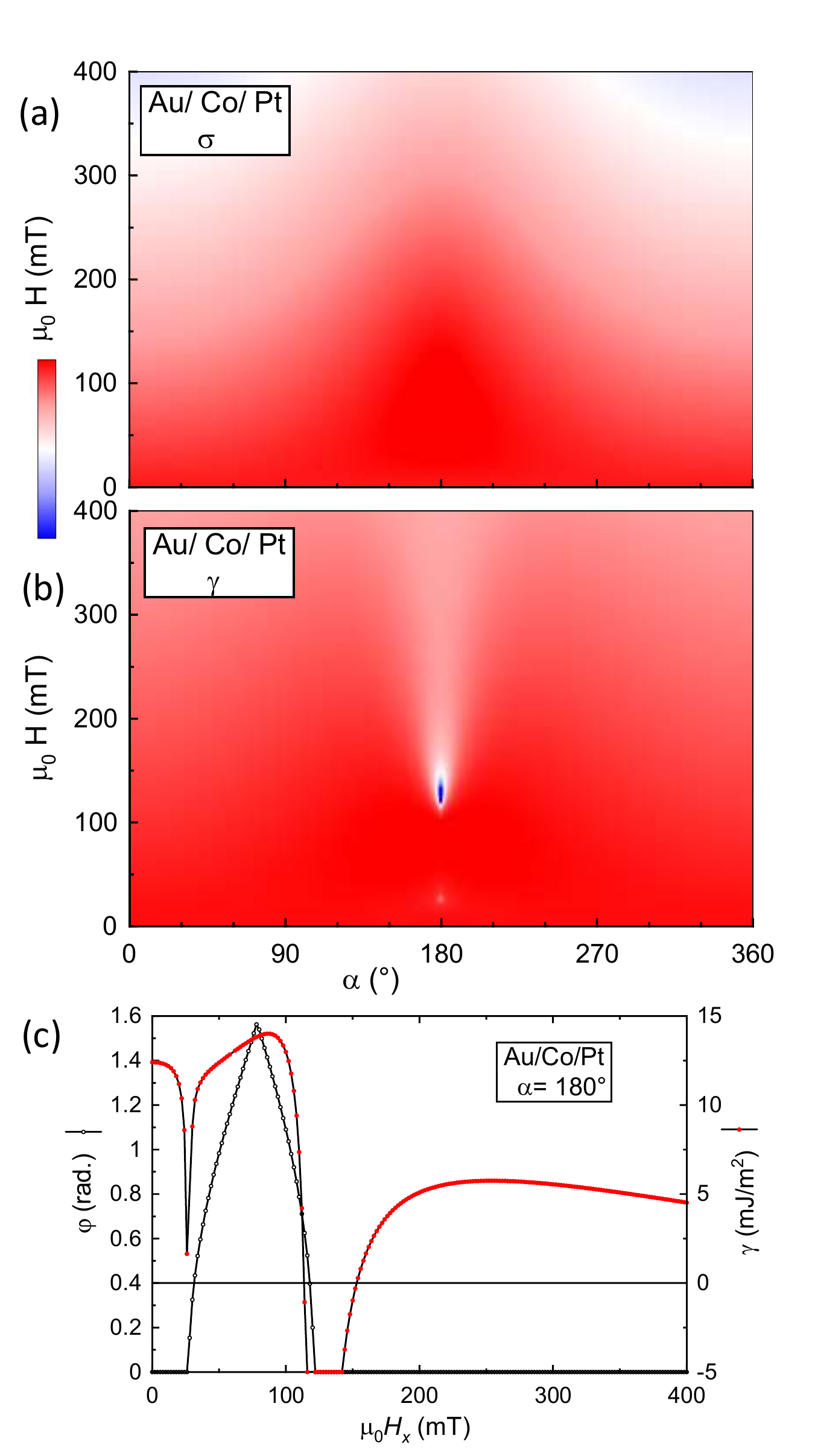}
\caption{Computed domain wall surface energy $\sigma$ (a) and tension $\gamma$ (b) for
the Au/Co/Pt sample, by the small circle model incorporating the additional demagnetizing
energy.
Note that such noiseless maps, containing 360x200 pixels, would have required very long numerical
micromagnetics calculation times, whereas a few seconds suffice to produce them using the small
circle model, including the magnetostatic correction.
The color scale spans the values 0 to +14~mJ/m$^2$ for (a), and -14 to +14~mJ/m$^2$ for
(b), values outside of these boundaries having been clipped.
Panel (c) compares a cut of (b) through $\alpha=180^\circ$ to the variation of the
small circle cut angle $\varphi$.
\label{fig:ACP-cartes}
}
\end{figure}
\begin{figure}
\includegraphics[width= 8 cm]{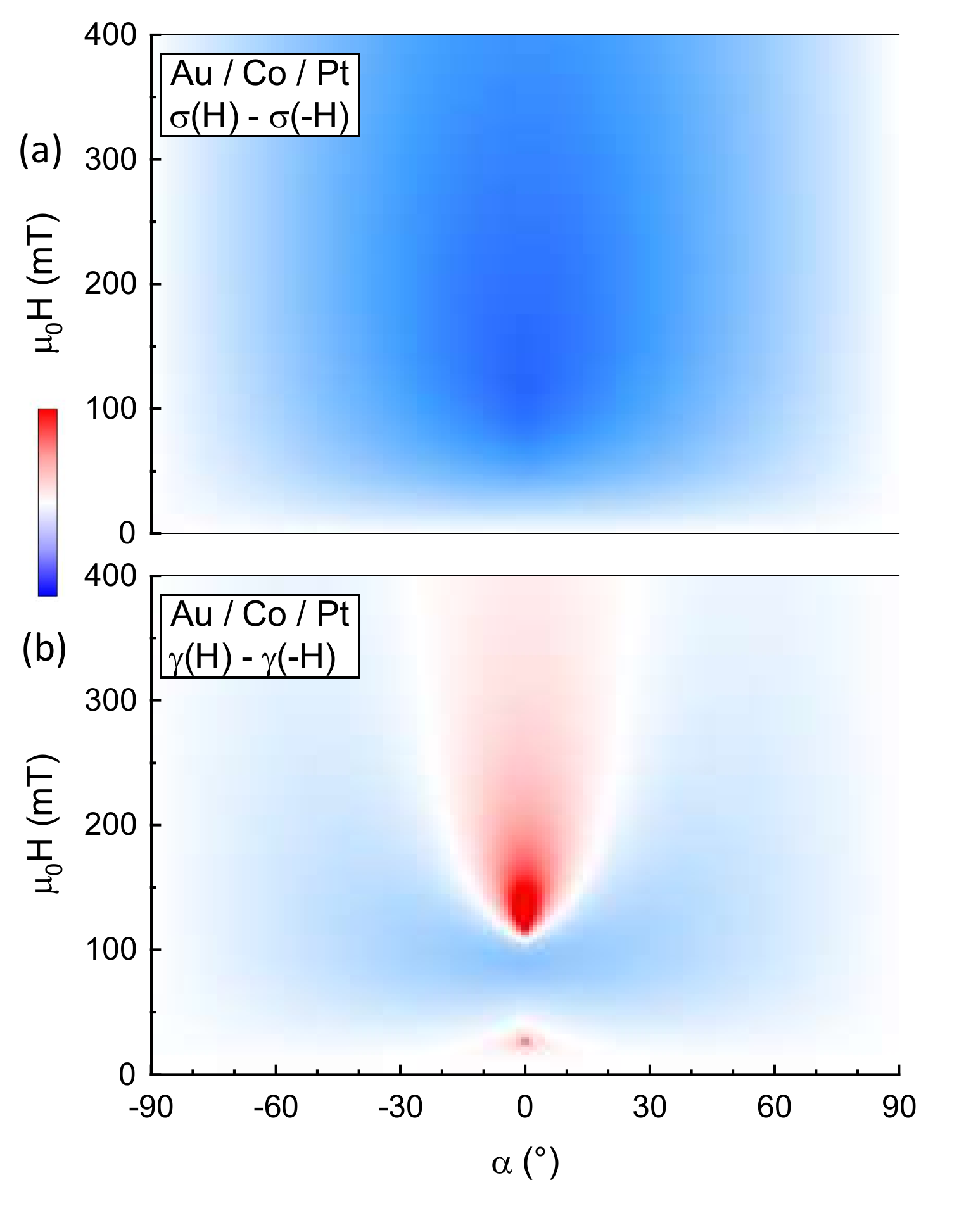}
\caption{Computed differences of domain wall surface energy $\sigma$ (a) and tension $\gamma$ (b) for
the Au/Co/Pt sample, between positive and negative fields $\epsilon(\alpha,H) - \epsilon(\alpha,-H)$.
The maps are derived from the data of Fig.~\ref{fig:ACP-cartes}.
The color scale spans values -5 to +5~mJ/m$^2$ for (a), and -10 to +10~mJ/m$^2$ for
(b), values outside of these boundaries having been clipped.
\label{fig:ACP-cartes-diff}
}
\end{figure}

The comparison of the two models, for the case of the Au/Co/Pt sample, is shown in 
Fig.~\ref{fig:compa-gamma}.
The quantitative agreement is close to perfect.
The two quantities $\sigma$ and $\gamma$ show a strikingly different behavior, even if the applied 
fields are all above the DMI field: whereas the energy $\sigma$ monotonously varies with angle, by a 
small amount, and monotonously decreases as more field is applied, the tension $\gamma$ shows a
marked decrease at intermediate fields, around the antiparallel orientation of the applied 
field with respect to the DMI field.
This difference comes from the strong sensitivity of $\gamma$ to the $\sigma(\alpha)$ variation.
Note for example that, if $\sigma= A \cos \alpha$ then $\gamma=0$.

In order to see better the difference between domain wall tension and energy, the alternative
plot where field is varied, for given values of the angle, is shown in Fig.~\ref{fig:compa-gamma-2}.

Now that the quantitative accuracy of the small circle Ansatz has been demonstrated, the model
can be used to obtain detailed predictions.
As an example, Fig.~\ref{fig:ACP-cartes} shows the computed color-coded maps of the domain
wall surface energy $\sigma$ and tension $\gamma$, for the case of the Au/Co/Pt sample.
Whereas the $\sigma$-map (a) shows the expected larger energy when field is antiparallel to the
DMI field (here, at $\alpha= 180^\circ$), the $\gamma$-map (b) shows an energy reduction around
that orientation, that depends strongly on the applied field.
The cut at $\alpha= 180^\circ$ (c) compares the variation of domain wall surface tension with that
of the cut angle $\varphi$ of the small circle, revealing that the deep troughs of $\gamma$ occur
when the domain wall magnetization reorients out of the N{\'{e}}el state.
The corresponding maps (not shown) for the symmetrical Pt/Co/Pt sample only show troughs in $\gamma$
at $\alpha \approx 0, 180, 360^\circ$ and $\mu_0 H \approx 26$~mT.

From these energy maps, maps of energy differences can be constructed, by comparing for the same angle
the results for positive and negative fields.
These are shown in Fig.~\ref{fig:ACP-cartes-diff}, for the surface energy and for the surface tension.
For Au/Co/Pt which has a positive DMI, and as up-down walls are considered, one expects that
$\sigma(H) < \sigma(-H)$ for positive fields. 
This is indeed obtained [Fig.~\ref{fig:ACP-cartes-diff}(a)].
However, and as remarked for the large domain wall anisotropy sample, the domain wall tension 
difference changes sign twice as field is increased [Fig.~\ref{fig:ACP-cartes-diff}(b)].
Therefore, if domain wall tension $\gamma$ alone were determining the domain wall velocity, one
would expect that the asymmetry of a circular-shape domain expanding in the presence of an
in-plane field would reverse twice, giving a sign in accord with that of the energy difference
only at intermediate fields.
This directly relates to experimental observations \cite{Vanatka15,Lau16}, as qualitatively 
explained earlier \cite{Pellegren17}.

\section{Conclusion}

We have developped a semi-analytical model for the one-dimensional domain wall structure in ultrathin 
films with perpendicular magnetization, in the presence of arbitrary in-plane fields, in orientation and
magnitude.
The model is based on the `small circle' Ansatz introduced by A. Hubert. 
It is semi-analytical, as an analytic expression needs to be minimized versus one variable,
the `cut angle' of the small circle.

The model has been compared to the simplest model of the situation, in which only the orientation of
the domain wall magnetic moment is allowed to vary. 
Clear differences have been observed, that increase as the in-plane field becomes larger.
The model ouputs have also been compared to numerical micromagnetic calculations, for three samples
having the Au/Co/Pt generic structure, the parameters of which coming from experiments.
A very good quantitative agreement has been obtained, with some systematic differences having been
uncovered, linked to the magnetostatic energy.
A correction to the domain wall energy has been worked out, which leads to much closer values.
For the domain wall width however, this is not generally possible as the model computes the full
structure of the domain wall.

The model provides the energy and the complete profile of the domain wall, which allows computing the 
various domain wall widths that are relevant for its statics (Hubert domain wall width), or dynamics (Thiele
domain wall width), or any other quantity dependent on the domain wall profile.
The complete freedom on the in-plane field allows computing the domain wall surface tension, whose key role 
has been recently uncovered, with no fear from artefacts due to too restricted energy calculation
hypotheses.
In particular, the occurence of zero tension regions (in the field-angle space) has been confirmed, meaning
that the one-dimensional picture breaks down there, and domain wall faceting occurs.
To illustrate the power of semi-analytical means, maps of the domain wall properties as a function
of the magnitude and angle of the in-plane field are shown.

It is hoped that the refined calculation of the domain wall properties developed in this work will be useful
in constructing a domain wall creep theory which fully incorporates the presence of the in-plane field (a
first step being Ref.~\cite{Hartmann19}), quantitatively explaining the surprising results of some 
experiments \cite{Lavrijsen15,Vanatka15,Lau16}.
The extension of this methodology to the fast domain wall dynamics should also be investigated.

\section{Acknowledgements}

This work was supported by Agence Nationale de la Recherche, projects
ANR-14-CE26-0012 (ULTRASKY) and ANR-17-CE24-0025 (TOPSKY).
We thank Jacques Miltat for a critical reading of the text.

\section{Appendix A: Analytical correction to the domain wall magnetostatic energy}

In the limit of a magnetization uniform across the sample thickness ($t$), which applies to
ultrathin films, the demagnetizing energy $E_\mathrm{d}$ of a domain wall with a Bloch profile (domain
wall width parameter $\Delta$) can be analytically calculated, by going to Fourier space.
This energy diverges, but the difference between two values of $\Delta$ is finite.
One obtains $E_\mathrm{d}(\Delta=0)-E_\mathrm{d}(\Delta)= (\mu_0 M_\mathrm{s}^2/2) t I\left( \Delta/t \right)$,
where the integral $I$ reads
\begin{equation}
I(p) = \int_0^{+\infty} {\frac{1-e^{-x}}{x} \left[ \frac{4}{\pi x^2} - 
\frac{\pi p^2}{\sinh^2(\pi p x /2)} \right] dx}.
\end{equation}

\begin{figure}[t]
\includegraphics[width= 8 cm]{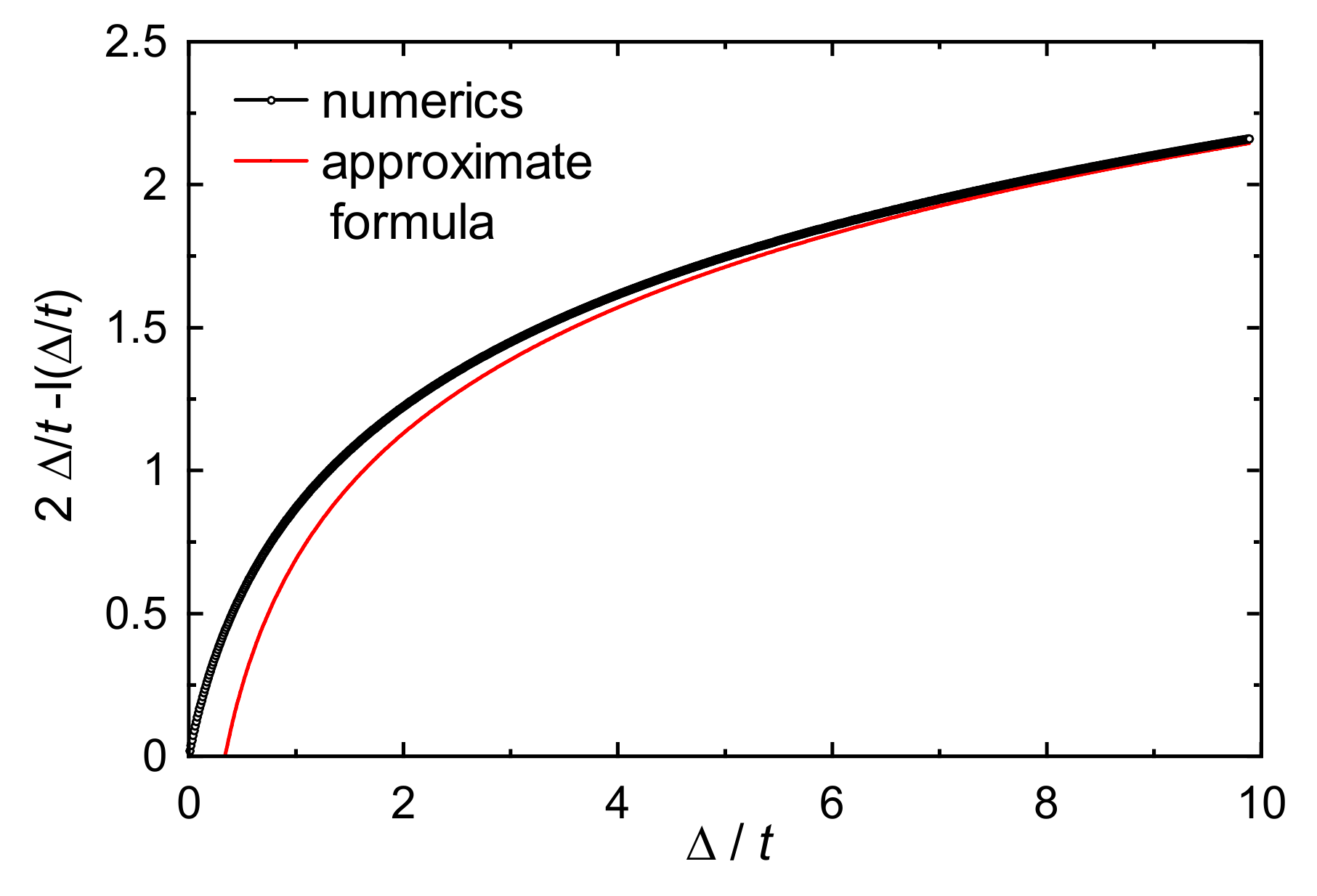}
\caption{Numerical evaluation of the function giving the demagnetizing energy correction $\sigma_\mathrm{d}$
to the domain wall energy.
The large parameter (i.e. $\Delta \gg t$) asymptotic behavior $0.69 + 0.635 \ln(\Delta / t)$ is also drawn.
\label{fig:sigma_d}
}
\end{figure}

On the other hand, the assumption of a local demagnetizing field leads to
$E_\mathrm{d}(\Delta=0)-E_\mathrm{d}(\Delta)= (\mu_0 M_\mathrm{s}^2/2) \left( 2 \Delta \right)$.
Therefore, the small circle domain wall energy $\sigma$ should be corrected by adding to it
the quantity $\sigma_\mathrm{d} = (\mu_0 M_\mathrm{s}^2/2)  \left[ 2 \Delta - t I(\Delta / t) \right]$.
The function involved is plotted in Fig.~\ref{fig:sigma_d}.
Under an in-plane field, $\sigma_\mathrm{d}$ should be multiplied by
$m_z^2=(1-h^2)$.

This magnetostatic domain wall energy correction $\sigma_\mathrm{d}$ can be used to predict the domain
wall width more accurately, in the case of zero applied field where the domain wall profile is known.
Taking this domain wall width parameter $\Delta$ as a variable, one has to minimize the
total energy of the domain wall, obtained by integrating the terms of Eq.~(\ref{eq:lesE}), which reads
\begin{equation}
\sigma(\Delta) = \frac{2A}{\Delta} + 2 K_0 \Delta + \frac{\mu_0 M_\mathrm{s}^2}{2} t
\left[ 0.69 + 0.635 \ln(\frac{\Delta}{t}) \right]
\end{equation}
(the Bloch-N{\'{e}}el magnetosatic cost, using the first order approximation for the demagnetizing
factor of the N{\'{e}}el wall in an ultrathin film \cite{Tarasenko98}, is independent of the 
domain wall width parameter).
This results in
\begin{equation}
\Delta = \sqrt{\Delta_0^2 + \left( \frac{0.635 t}{4 Q_0} \right)^2}-\frac{0.635 t}{4 Q_0},
\end{equation}
where $Q_0 = 2 K_0 / (\mu_0 M_\mathrm{s}^2)$ is the quality factor of the sample.
As an example, for the Au/Co/Pt sample, one finds $\Delta= 4.385$~nm, much closer to the numerical
value.
This calculation shows again the origin of the discrepancy between the small circle and the full
micromagnetics.



\begin{thebibliography}{10}
\bibitem{note-PC}
The small circle model was already employed in a previous publication by some
  of us \cite{Pizzini14}, but not described in depth and without incorporating
  the domain wall internal magnetostatic energy, as it played a minor role in
  that case.
\bibitem{note-sigma}
For the numerical simulations, three configurations had to be considered: (i)
  the minimized configuration with a domain wall in the center, (ii) the
  minimized configuration starting from the all-up state (as well as from the
  all-down one), (iii) the zero-thickness domain wall configuration, produced
  by pasting the left half of the all-up to the right half of the all-down
  minimized configurations. For configuration (iii), no evolution was performed
  and only the magnetostatic energy was evaluated. The domain wall energy was
  obtained as $\sigma = L \left[ \mathcal{E}(\mathrm{i}) -
  \mathcal{E}(\mathrm{ii}) + \mathcal{E}_\mathrm{dem}(\mathrm{ii}) -
  \mathcal{E}_\mathrm{dem}(\mathrm{iii}) \right]$, where $L$ is the width of
  the central region (400~nm here) and $\mathcal{E}$ resp.
  $\mathcal{E}_\mathrm{dem}$ are total resp. magnetostatic average energy
  densities computed over that region. The last two terms in the equation allow
  removing the magnetostatic energy gained when demagnetizing a sample by
  having it split into two domains. The first difference is necessary because
  in MuMax3 a uniform configuration magnetized along the easy axis has a large
  (negative) energy. The domain wall energy is thus obtained by differences
  from large terms, so that accuracy of these terms is very important.
\end{thebibliography}
\end{document}